\documentclass[aps]{revtex4}
\usepackage{amsfonts}
\usepackage{amsmath}
\usepackage{graphicx}
\usepackage{subfigure}
\usepackage{amssymb,epsf}
\usepackage{float}
\usepackage{color}

\begin{document}

\title{Dynamical and Thermal Stabilities of Nonlinearly Charged AdS Black Holes}
\author{S. N. Sajadi$^{1}$ \footnote{
email address: naseh.sajadi@gmail.com}, N. Riazi$^{1}$ \footnote{
email address: n\_riazi@sbu.ac.ir} and S. H. Hendi $^{2,3}$
\footnote{ email address: hendi@shirazu.ac.ir}}
\affiliation{$^{1}$ Department of Physics, Shahid Beheshti
University, G.C., Evin, Tehran 19839,  Iran\\
$^{2}$ Physics Department and Biruni Observatory, College of
Sciences, Shiraz
University, Shiraz 71454, Iran\\
$^{3}$ Research Institute for Astronomy and Astrophysics of
Maragha (RIAAM), P.O. Box 55134-441, Maragha, Iran}

\begin{abstract}
In this paper, we study an extended phase space thermodynamics of
a nonlinearly charged AdS black hole. We examine both the local
and global stabilities, and possible phase transition of the black
hole solutions. Finally, we compute quasi-normal modes via scalar
perturbations and compare the obtained results with those of
RN-AdS black hole.
\end{abstract}

\maketitle

\section{Introduction}
Nonlinear electrodynamics was first proposed by Born and Infeld in
order to remove the central singularity of the point-like charges
and obtain finite energy solutions for particles by extending
Maxwell theory \cite{Born:1934ji}. Later, Plebanski and
Perzanowski extended the model and presented other examples of
nonlinear electrodynamics Lagrangians in the framework of special
relativity \cite{Plebanski:1994je}. Recently, studying various
models of nonlinear electrodynamics has been under active
investigation, mainly because these theories appear as effective
theories at low energy limits of string theory
\cite{Seiberg:1999vs}. In the framework of AdS-CFT, nonlinear
electrodynamics has been used to obtain solutions describing
baryon configurations which are consistent with confinement
\cite{Aharony:1999ti}. In other frameworks, in order to remove
curvature singularities various important results have been
obtained. For example, in cosmological models, one can use the
nonlinear electrodynamics for explaining the inflationary epoch
and the late-time accelerated expansion of the universe
\cite{GarciaSalcedo:2000eb},\cite{Novello:2003kh}. In the field of
black holes, different classes of regular black hole solutions in
general relativity coupled to nonlinear electrodynamics have been
found, in which the nonlinear electrodynamics is a source of field
equations satisfying the weak energy condition, and recovering the
Maxwell theory in the weak field limit \cite{29}-\cite{40}. There
are some nonlinear electrodynamics Lagrangians that can not remove
singularity from the black holes, but they generalize the RN
solution. For example, Born-Infeld Lagrangian gives rise to a
singular spherically symmetric black hole and other solutions
\cite{Wiltshire:1988uq}-\cite{Soleng:1995kn}. The progress in this
direction and demand to investigate various aspects of nonlinear
models are the main motivations for the present study. In this
paper, we study an interesting class of nonlinear electrodynamics,
in which regularizes the metric function. Regarding such class of
nonlinear electrodynamics, we find that although this model can
regularize the metric function, the curvature scalars diverge at
the origin.

Thermodynamic properties of black holes were reinforced by the
discovery of Hawking radiation \cite{1}. Bekenstein considered the
concept of entropy for black holes and made it quantitative,
namely the area law, $S=A/4$ \cite{2}. However, there are some
differences between black hole thermodynamics and conventional
thermodynamics, such as the black hole entropy, which is
proportional to the horizon area and not volume, and the heat
capacity of black holes which might be negative.

Thermodynamic properties of AdS black holes are more interesting
due to some reasons: one of them comes from the main work of
Hawking and Page, who discovered a first order phase transition
between the Schwarzschild-AdS black hole and thermal AdS space
\cite{3}. This phenomenon, known as the Hawking-Page phase
transition, is explained as the gravitational dual of the QCD
confinement/deconfinement transition \cite{4,5}. Another important
reason in the thermodynamics of AdS black holes was the discovery
of phase transitions similar to Van der Waals liquid/gas phase
transitions in the Reissner-Nordstr${\rm \ddot{o}} $m/anti-de
Sitter (RN-AdS) black holes by Chambline et al.\cite{6,7}. They
studied the RN-AdS black holes in canonical ensemble and
discovered a first order phase transition between small and large
black holes. Another motivation of considering AdS black holes is
due to the AdS-CFT correspondence \cite{8}. This duality has been
recently used to study the behavior of quark-gluon plasmas and the
qualitative description of various condensed matter phenomena.
According to the AdS/CFT correspondence \cite{jm}, a large static
black hole in asymptotically AdS spacetime corresponds to a
thermal state in the CFT living on the boundary.

Actually, the interests in studies of thermodynamics of the AdS
black holes is partly due to the rich structures found by treating
the cosmological constant as a thermodynamic variable. In the
presence of a varying cosmological constant, the first law of
black hole thermodynamics becomes consistent with the Smarr
relation. The first law of black holes is modified by including a
$V dP$ term where the pressure $P$ is given by
$-\frac{\Lambda}{8\pi}$. In this framework, the mass of the black
hole $M$ is considered as the enthalpy of the system instead of
the internal energy \cite{9,10}. One of the first works to explore
the extended phase space thermodynamics in AdS black holes was
written by Kubiznak and Mann who showed the existence of a certain
phase transition in the phase space of the Schwarzschild/AdS black
hole \cite{11}. Similar critical behavior is found in the
spacetimes of a rotating AdS black hole and a higher dimensional
RN-AdS black hole \cite{12}. The same qualitative properties are
also found in the AdS black hole spacetime with the Born-Infeld
electrodynamics \cite{13,14}, with the power-Maxwell field and
with Gauss-Bonnet correction \cite{15,16}. There are many other
works related to this concept including \cite{17}-\cite{28}.

In this paper, we introduce a new charged AdS black hole solution,
in the presence of a nonlinear electrodynamics. We then perform a
detailed analysis of the thermodynamics of the obtained BH
solution. Also, we study the behavior of a scalar field outside
the black hole horizon by computing the complex frequencies
associated with quasinormal modes. These modes are the resonant,
non-radial perturbations of black holes which are the result of
external perturbations of spacetime. Quasinormal modes are
important for some reasons. First, because information about the
black hole is encoded in there. By detection of the quasi normal
modes, one can obtain precise information related to the mass,
charge, and other global parameters of the black hole. Another
reason is that complex frequencies can provide information about
dynamical stability of the black hole. Indeed, the real part of
frequencies determines the oscillation frequency and the imaginary
part determines the rate at which each mode is damped. The final
reason comes from the fact that the quasinormal frequencies of AdS
black holes have an interpretation in the dual conformal field
theory \cite{db}. According to this correspondence, a large black
hole in AdS corresponds to a thermal state in the CFT. Perturbing
the black hole, corresponds to perturbing the thermal state and
the decay of the perturbation describes the return to thermal
equilibrium. Therefore, the time scale for the decay of the black
hole perturbation, which is given by the imaginary part of its
QNM, corresponds to the timescale to reach thermal equilibrium in
the strongly coupled CFT \cite{db},\cite{gth},\cite{gth2},\cite{
bwe}.

The paper is organized as follows: in section \ref{base}, the
basic equations are introduced. Section \ref{sec3} is devoted to
the calculation of conserved and thermal quantities and
investigation of the first law of thermodynamics. Subsequently,
thermal stability and phase transition are discussed. In section
\ref{qnm}, we compute quasi-normal modes and finally, our
concluding remarks are given in section \ref{sec4}.
\section{Basic Equations}\label{base}
The 4-dimensional action governing charged black holes in the presence of a nonlinear electromagnetic field is given by \cite{29}-\cite{46}
\begin{equation}\label{eqI}
I=\frac{1}{4\pi}\int d^{4} x \sqrt{-g}\left[ \frac{1}{4
}(R-2\Lambda) -L(P)\right],
\end{equation}
where $ R $ is the Ricci scalar and $ \Lambda=-\frac{3}{b^{2}} $
is the cosmological constant, in which $ b $ denotes the radius of
AdS space. $ L(P) $ is a function of $ P
\equiv\frac{1}{4}P_{\mu\nu}P^{\mu\nu }$ and $ P_{\mu\nu} $ is an
anti{symmetric tensor related to the Faraday tensor $ F_{\mu \nu}
$ according to $ P_{\mu \nu}= L_{F}F_{\mu \nu}  $ and $
L_{F}\equiv\frac{\partial L}{\partial F} $ ($ F\cong F_{\mu
\nu}F^{\mu \nu} $ is the Maxwell invariant). Applying the
variational principle to the action (\ref{eqI}), one can show that
the field equations are given by \cite{29}-\cite{46}
\begin{equation}\label{gg}
G^{\nu}_{\mu}+\Lambda
\delta^{\nu}_{\mu}=\frac{1}{2}\delta^{\nu}_{\mu}L-2L_{P}P_{\mu
\lambda}P^{\nu \lambda},
\end{equation}
\begin{equation}\label{nn}
\nabla_{\mu}P^{\alpha\mu}=0,
\end{equation}
where in the above equations, $ G_{\mu \nu} $ is the Einstein
tensor and $ L_{P}\equiv\frac{\partial L}{\partial P} $
\cite{29}-\cite{46}. By integrating equation (\ref{nn}) with the
assumption of spherical symmetry, we obtain the static solution
\begin{equation}
P_{\mu
\nu}=2\delta^{t}_{[\mu}\delta^{r}_{\nu]}\frac{Q}{r^{2}}\hspace{0.5cm}or
\hspace{0.5cm} P=-\frac{Q^{2}}{2r^{4}}
\end{equation}
where $ Q $ is an integration constant related to the electric charge.
Here, we assume that the Lagrangian is as follows \cite{30}-\cite{32}:
\begin{equation}\label{lagr}
L=\frac{8\alpha^{3}P(6\alpha(-2P)^{\frac{1}{4}}+\sqrt{\beta}-2\alpha
\beta
(-2P)^{\frac{3}{4}})}{\sqrt{\beta}(2\alpha+\beta^{\frac{3}{2}}(-2P)^{\frac{1}{4}}+2\alpha
\beta(-2P)^{\frac{1}{2}})^{3}}.
\end{equation}
In the weak field limit ($ P\ll 1 $) the EM Lagrangian reduces to
\begin{equation}
 L \approx P+\mathcal{O}(P^{\frac{5}{4}})
\end{equation}
With the nonlinear electric field as source, the gravitational field is described by the metric
\begin{equation}\label{me}
ds^{2}=-f(r)dt^{2}+f(r)^{-1}dr^{2}+r^{2}d\Omega^{2}
\hspace{0.5cm},\hspace{0.5cm} f(r)=1-\frac{\Lambda}{3}
r^{2}-\frac{2\alpha r}{r^{2}+\frac{\beta^{2}}{2\alpha}
r+\beta^{2}}.
\end{equation}
Indeed, one can put the metric (\ref{me}) into equation (\ref{gg}) and solve for $ L(P) $, to obtain the Lagrangian (\ref{lagr}). So, this Lagrangian is appropriate only for the present black hole and by using this Lagrangian one can not obtain regular black holes such Bardeen and other regular black holes.
The particular form of the Lagrangian (\ref{lagr}) thus leads to the interesting solution (\ref{me}) which contrary to the RN solution of Einstein-Maxwell is more usual. System, behaves regularly as $ r\rightarrow 0 $.
The associated electric field $ E $ is given by
\begin{equation}\label{eqel}
E=\frac{2\alpha^{2}\beta r(4\alpha r^{3}+(20\alpha^{2}+6\alpha
\beta^{2}+\beta^{2})r^{2}+4\alpha^{2}\beta^{2})}{(2\alpha
r^{2}+\beta^{2}r+2\alpha \beta^{2})^{3}}.
\end{equation}
The asymptotic behavior of the solution as $ r\rightarrow \infty $ reads
\begin{equation}
-g_{tt}\approx1-\frac{2\alpha}{r}+\frac{\beta^{2}}{r^{2}}-\frac{\Lambda}{3}
r^{2} \hspace{0.5cm},\hspace{0.5cm}
E\approx\frac{\beta}{r^{2}}+\mathcal{O}(\frac{1}{r^{3}}).
\end{equation}
In comparison to the RN-AdS metric and coulomb law, we observe that $ M $ and $ Q $ are related to mass and charge according to
\begin{equation}
M=\alpha\hspace{0.5cm},\hspace{0.5cm}Q=\beta,
\end{equation}
where $ M $ is ADM mass and $ Q $ is electric charge of the solution. Note that in contrast to the Einstein-Maxwell system, $M$ and $Q$ are not constants of integration, but rather are parameters present in EM Lagrangian (\ref{lagr}).
We plot diagrams for finding the possible roots of the metric function which correspond to horizons (Fig. \ref{fig1}). We find from this figure that our solution can represent a black hole, by suitable choices of parameters. For $ Q=0.827M $, the horizons degenerate to a single one, corresponding to an extreme black hole. Therefore, there is a minimal mass
\begin{equation}
M_{min}=1.21Q.
\end{equation}
 Below this mass, we would have a mild naked singularity.
\begin{center}
\begin{figure}[H] \epsfxsize=8cm \epsffile{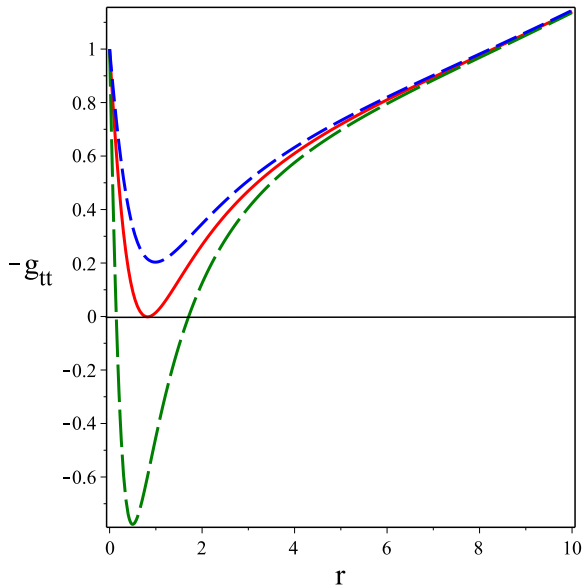}\vspace{0.1cm}
\caption{ The behavior of $ -g_{tt} $ in terms of $ r $ for $
Q=0.5 $ with no horizons ({long dashed line}), $Q=0.827$ extremal
({solid line }) and $Q=1$ with two horizons ({dashed line}) and
$\Lambda=-0.01$, $ M=1 $ (all three cases).} \label{fig1}
\end{figure}
\end{center}
It is worth noting that the metric (\ref{me}) is well behaved when $ r $ goes to zero. The electric field vanishes as $r \rightarrow 0$ and $ r\rightarrow \infty $ and achieves its maximum $ E_{c} $ at $ r_{c} $ (see figure \ref{fig21}). In order to
get information the nature of $ r_{c} $, see \cite{snsr} and the reference therein.
\begin{center}
\begin{figure}[H] \epsfxsize=8cm \epsffile{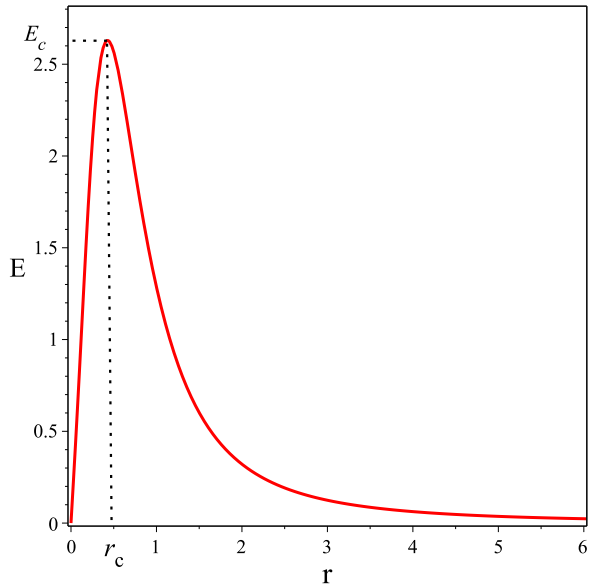}\vspace{0.1cm}\caption{\label{fig21} \small
 The behavior of the electric field in terms of $r$ for $M = 1, Q = 0.5$.}
\end{figure}
\end{center}
The limit of Ricci and Kretschmann scalar $ K_{1}=R_{abcd}R^{abcd} $  for metric (\ref{me}) when $ r \rightarrow 0 $ are
\begin{equation}
\lim_{r=0}  K_{1}=\frac{32 M^{2}}{Q^{4}
r^{2}},\hspace{0.5cm}\lim_{r=0} R=\frac{12M}{Q^{2} r}
\end{equation}
As one can see, when $ r $ goes to zero the Kretschmann scalar diverges as $ \frac{1}{r^{2}} $, for all values of $ Q $ and $ M $. For comparison, $K_{1}\propto \frac{1}{r^{6}}$ for the RN metric. \\
To derive the global structure of the metric, one can construct the Penrose diagrams. The Penrose diagram of the present metric is similar to the Reissner-Nordstr${\rm \ddot{o}}$m-AdS metric. In the case of non-extreme black hole solution $ Q<0.827M $, we can split the space-time into three regions; $ I: r>r_{+} $, $ II:r_{-}<r<r_{+} $ and $ III: 0\leq r<r_{-}$. In the extreme black hole case, $ Q=0.827M $, there arise two regions: $ r>r_{c} $  and $ 0<r<r_{c} $. For $ Q>0.827M $, there are no horizons.\\
\section{Thermodynamic Properties}\label{sec3}
In order to investigate the thermodynamic stability of the black hole in extended phase space, we need to obtain some relevant thermodynamic quantities.
The black hole temperature for non-extremal case with two horizons $ r_{\pm} $ reads
\begin{equation}\label{eq15}
T=\frac{f^{'}(r_{+})}{4\pi}={\frac { \left( -24Q^{2}+24r_{+}^{2}
\right) M^{3}-8 \Lambda r_{+} \left(Q^{2}+r_{+}^{2} \right)
^{2}M^{2}-8r_{+}^{2}Q^{2 }\Lambda \left( Q^{2}+r_{+}^{2} \right)
M-2Q^{4}r_{+}^{3}\Lambda}{3
 \left( 2Q^{2}M+2r_{+}^{2}M+{Q}^{2}r_{+} \right) ^{2}}},
\end{equation}
 where $ r_{+} $ is the event horizon which is defined by the larger root of $f(r_{+}) = 0$. Contrary to RN black hole, as $ r_{+} $ vanishes, the temperature does not diverge but has a finite value.
The black hole entropy $ S $ can be calculated via the area law
\begin{equation}\label{eq16}
 S=\frac{A}{4}=\pi r_{+}^{2}
\end{equation}
and by using equation (\ref{eqel}), the electrostatic potential $ \phi $ is obtained as
\begin{equation}\label{eq17}
\phi =\int_{r_{+}}^{\infty}E dr={\frac
{\left(8{r_{+}}^{3}M+20{r_{+}}^{2}{M}^{2}+3{r_{+}}^{2}{Q}^{2}+12r_{+}M{Q}^{2}+12{M}^{2}{Q}^{2}
 \right) M{Q}^{2}}{ 2\left( 2{r_{+}}^{2}M+r_{+}{Q}^{2}+2
M{Q}^{2}\right)^{2}Q}}.
\end{equation}
The pressure $ P $ is defined as \cite{11}:
\begin{equation}
P=-\frac{\Lambda}{8 \pi},
\end{equation}
with the conjugate quantity which is volume:
\begin{equation}\label{eq19}
V=\frac{4}{3}\pi r^{3}_{+}.
\end{equation}
In equations (\ref{eq15}) and (\ref{eq17}), $ M $ is related to $ r_{+} $ and other BH parameters as
\begin{eqnarray}\label{eqm}
 M&=&\frac{-Q^{2}{r^{2}_{+}}\Lambda- r_{+}^{4}\Lambda+3 Q^{2}+3r_{+}^{2}}{12r_{+}}+\\
&&\frac{\left(
Q^{4}r_{+}^{4}\Lambda^{2}+2Q^{2}r_{+}^{6}\Lambda^{2}+
r_{+}^{8}\Lambda^{2}-6Q^{4}r_{+}^{2}\Lambda -24Q^{2}r_{+}^{4}
\Lambda -6r_{+}^{6}\Lambda
+9Q^{4}+54Q^{2}r_{+}^{2}+9r_{+}^{4}\right)^{\frac{1}{2}}}
{12r_{+}}.
\end{eqnarray}
For a black hole embedded in AdS spacetime and by employing the relation between the cosmological constant and thermodynamic pressure, the black hole mass $ M$ will be interpreted as the enthalpy $H=M$ \cite{10,11,24,bp,dks,dcz}.
So, the enthalpy in terms of thermodynamic quantities is given by
\begin{eqnarray}
H(S,P,Q)&=& \frac{8\pi{Q}^{2}SP +3\pi{Q}^{2} +3S+8{S}^{2}P}{12\sqrt {\pi S}}+\\
&&\frac{\sqrt {8 PS+3}{\left(8{\pi }^{2}PS{Q}^{4}+16\pi P
{S}^{2}{Q}^{2}+3{\pi }^{2}{Q}^{4}+8{S}^{3}P+18\pi{Q}^{2}
S+3{S}^{2}\right)^{\frac{1}{2}} }}{12\sqrt { \pi S }}.
\end{eqnarray}
The usual thermodynamic relations can now be used to determine the
temperature, electrostatic potential and volume
\cite{6}-\cite{28},
\begin{equation}\label{eq21}
T=\left(\frac{\partial H}{\partial S} \right)_{P Q},
\end{equation}
\begin{equation}\label{eq23}
 \phi=\left(\frac{\partial H}{\partial Q} \right)_{S P},
 \end{equation}
 and
\begin{equation}\label{eq22}
 V=\left(\frac{\partial H}{\partial P} \right)_{S Q}.
 \end{equation}

The Smarr relation is given by
 \begin{equation}
 \frac{H}{2}+PV-TS-\frac{Q \phi}{2}-\frac{1}{2}\int w dv=0,
 \end{equation}
 where
\begin{eqnarray}
\int w dv=\int_{r_{+}}^{\infty} \frac{T^{\mu}_{\mu}}{2}4\pi
r^{2}dr=\frac {128{M}^{2}{P}{Q}^{4}\pi
r_{+}^{3}+256{M}^{2}{P}{Q}^{2}\pi r_{+}^{5}+128{M}^{2}{P}\pi
r_{+}^{7}+128M{P}{Q}^{4}\pi r_{+}^{4}}{6\left(2M{Q}^{2}+2M
r_{+}^{2}+{Q}^{2}r_{+}\right)^{2}}\\+\frac{128M{P}{Q}^{2}\pi
r_{+}^{6}+32r_{+}^{5}{Q}^{4}\pi{P}+12{M}^{3}{Q}^{4}-12{M}^{3}{Q}^{2}
r_{+}^{2}+12{M}^{2}{Q}^{4}r_{+}+3M{Q}^{4}r_{+}^{2}}{6\left(2M{Q}^{2}+2M
r_{+}^{2}+{Q}^{2}r_{+}\right)^{2}},
\end{eqnarray}
this term comes from non vanishing trace of the energy-momentum
tensor \cite{Balart:2017dzt}. Calculations show that the intensive
quantities calculated by Eqs. (\ref{eq21}, \ref{eq23} and
\ref{eq22}) coincide with Eqs. (\ref{eq15}), (\ref{eq17}) and
(\ref{eq19}) respectively. Thus, these thermodynamic quantities
satisfy the first law of BH thermodynamics \cite{10}-\cite{24}
 \begin{equation}
dH=dM=TdS+\phi dQ+VdP.
\end{equation}}
\subsection{Specific Heats and Thermodynamic Stability}
Thermodynamic stability tells us how a system in thermodynamic equilibrium responds to fluctuations of energy, temperature and other thermodynamic parameters. We distinguish between global and local stability. In global stability we allow a system in equilibrium with a thermodynamic reservoir to exchange energy with the reservoir. The preferred phase of the system is the one that minimizes the Gibbs free energy \cite{15,26,27,28,shh,shh1}.
\begin{center}
\begin{figure}[H]\epsfxsize=8cm \epsffile{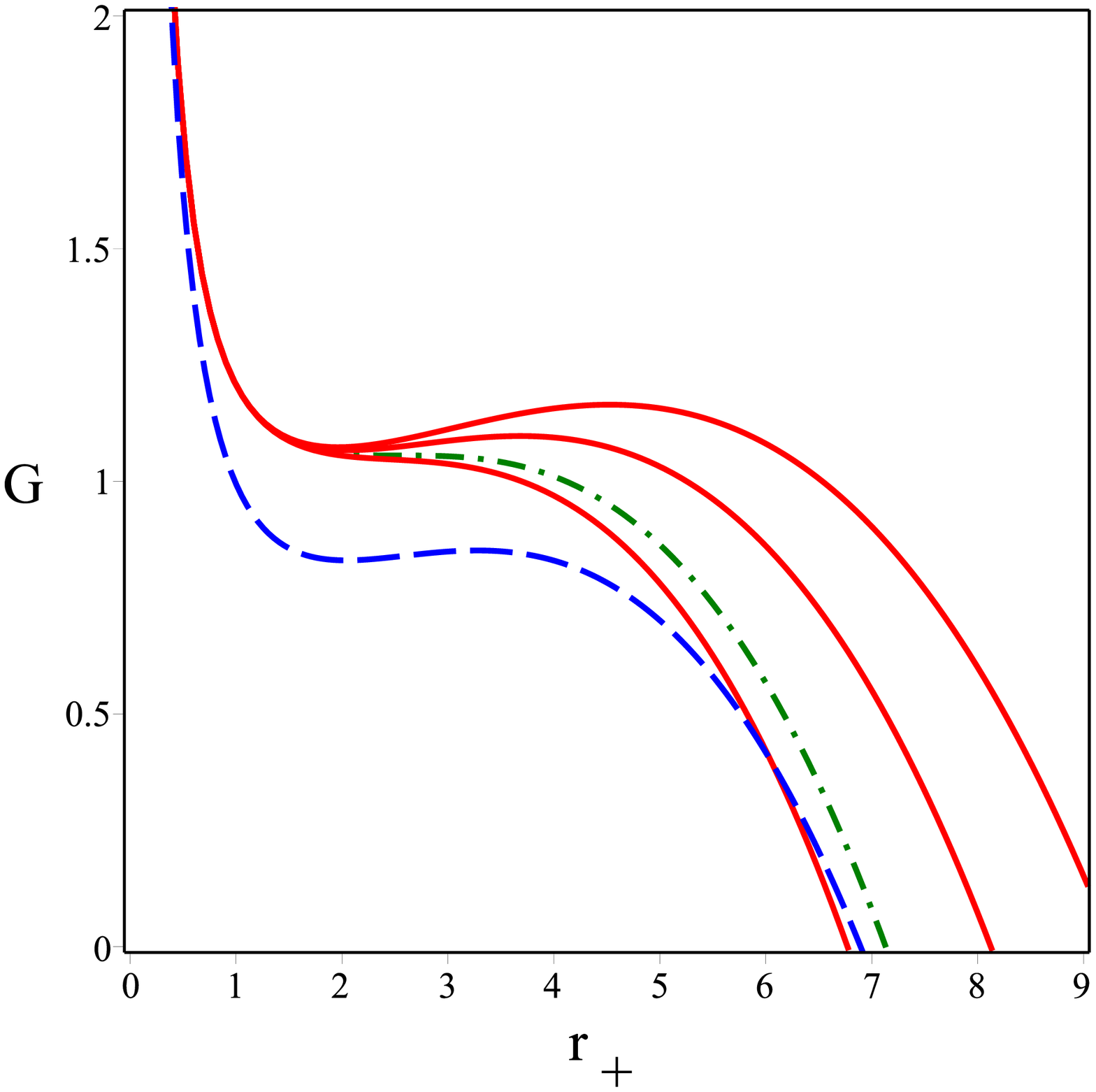}\vspace{0.1cm}\caption{\label{fig2}\small The behavior of $ G $
for our solution in terms of $ r_{+} $ for $
P=\textcolor{red}{0.003}, \textcolor{green}{0.002669},
\textcolor{red}{0.002}, \textcolor{red}{0.0015} $ (left to right)
and $ Q=1 $, and for RN-AdS metric ($ \textcolor{blue}{dashed}$
line) for $ P=\textcolor{blue}{0.002669}$, $ Q=1 $.}
\end{figure}
\end{center}
In order to investigate the global stability and phase transition of the system in canonical ensemble, we use the following expression for Gibbs free energy \cite{10,11,12,24}
\begin{equation}
G=H-T S.
\end{equation}
We find from Fig. \ref{fig2} that for constant pressure and electric charge, $G$ is a decreasing function for small/large event horizon, while it is an increasing function for intermediate $ r_{+} $. This behavior confirms that intermediate black holes are globally unstable. Large black holes have negative Gibbs free energy and therefore are more stable than small black holes. Also, one can see that for constant $ r_{+} $, for the larger pressure, the BH is more stable. Also, we compare the Gibbs free energy of the RN-AdS metric with the present solution. The Gibbs free energy could also have extremum by suitable choices of parameters. We shall discuss this issue in the next section when we study the phase transition.\\
Local stability is concerned with how the system responds to small changes in its thermodynamic parameters. In order to study the thermodynamic stability of the black holes with respect to small variations of the thermodynamic coordinates, one can investigate the behavior of the heat capacity. The positivity of the heat capacity ensures local stability. There are two different heat capacities associated with a system. $ C_{V} $: which measures the heat capacity when the heat is added to the system
keeping the volume constant and $ C_{P} $: which is used when the heat is added at constant pressure. Heat capacities can be calculated using the standard thermodynamic relations  \cite{shh,shh1}
\begin{equation}\label{eq26}
C_{V}=T\left( \frac{\partial S}{\partial
T}\right)_{V}\hspace{0.5cm}, \hspace{0.5cm}C_{P}=T\left(
\frac{\partial S}{\partial T}\right)_{P}.
\end{equation}
Using the expression for black hole entropy (Eq. \ref{eq16}) one can show that $ C_{V} $ vanishes.
$ C_{V} $ vanishes, while one can use Eq. (\ref{eq26}) to calculate $ C_{P} $, numerically, as its behavior is shown in Fig. \ref{fig3}.
\begin{center}
\begin{figure}[H] \epsfxsize=8cm \epsffile{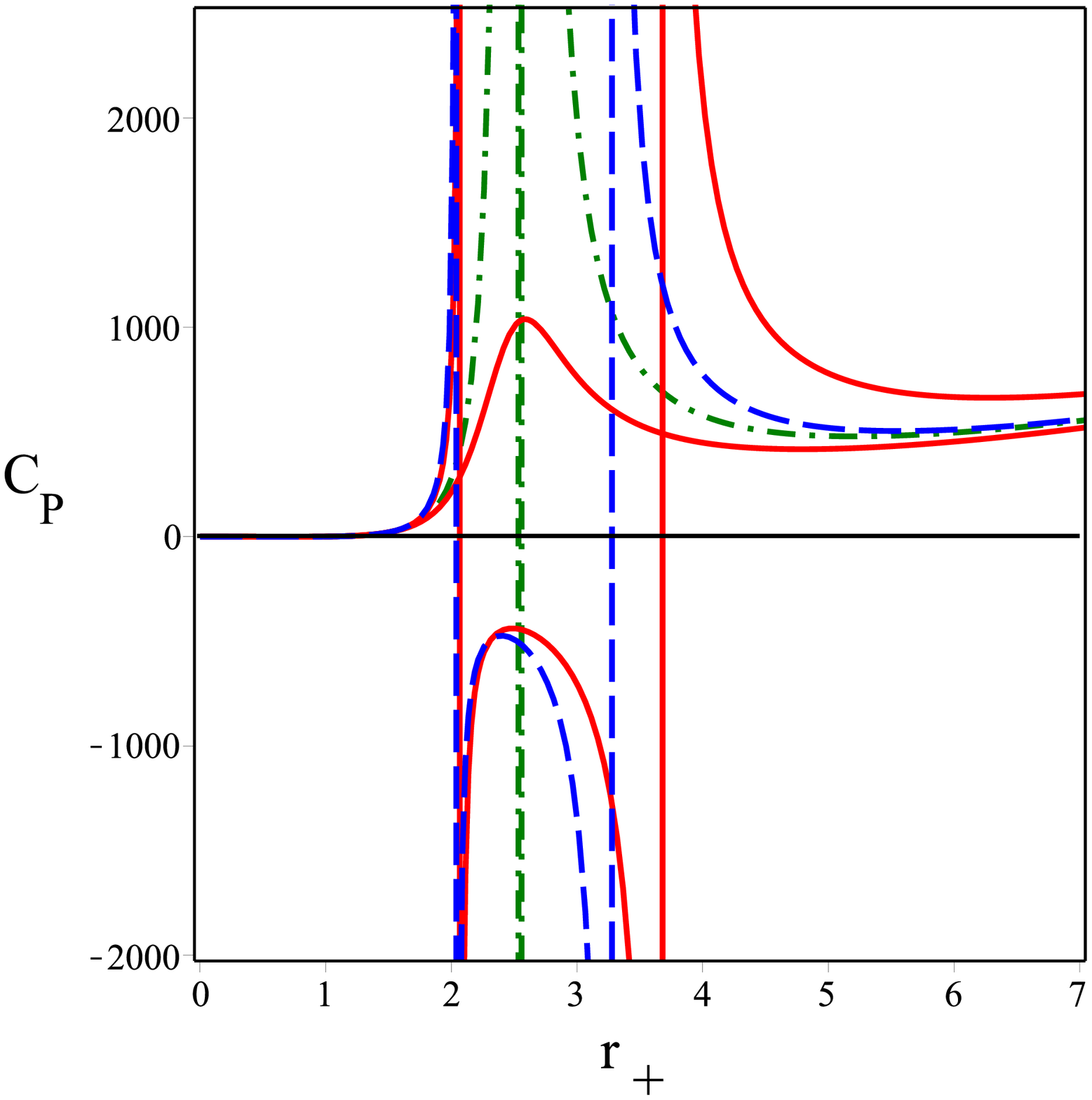}\vspace{0.1cm}\caption{\label{fig3} \small
The behavior of $ C_{P} $ in terms of $ r_{+} $ for $
P=\textcolor{red}{0.002}, \textcolor{green}{0.002669},
\textcolor{red}{0.003}$ (right to left) and $ Q=1 $, and for
RN-AdS metric ($ \textcolor{blue}{ dashed }$ line) and for $
P=\textcolor{blue}{0.002669}$, $ Q=1 $.}
\end{figure}
\end{center}
%\begin{figure}[H] \hspace{4cm}\includegraphics[width=8.cm]{CPP3}\vspace{0.1cm}\caption{\label{figg} \small
%The behavior of $ C_{P} $ in terms of $ r_{+} $ for $ P=5$, $ Q=1 $downward.}
%\end{figure}
%\end{center}
Regarding Fig. \ref{fig3}, we find that the heat capacity is positive for large black holes and partly positive for the smaller ones. It means that the large black hole is thermodynamically more stable (locally). For medium black holes, since the heat capacity is negative, the black hole is locally unstable, which is similar to RN-AdS black hole.
We shall discuss this issue in section \ref{sec4-2} when we consider the critical points in black hole phase diagram.
\subsection{Phase Transition}\label{sec4-2}
\begin{figure}[H]
\centering {\epsfxsize=8cm \epsffile{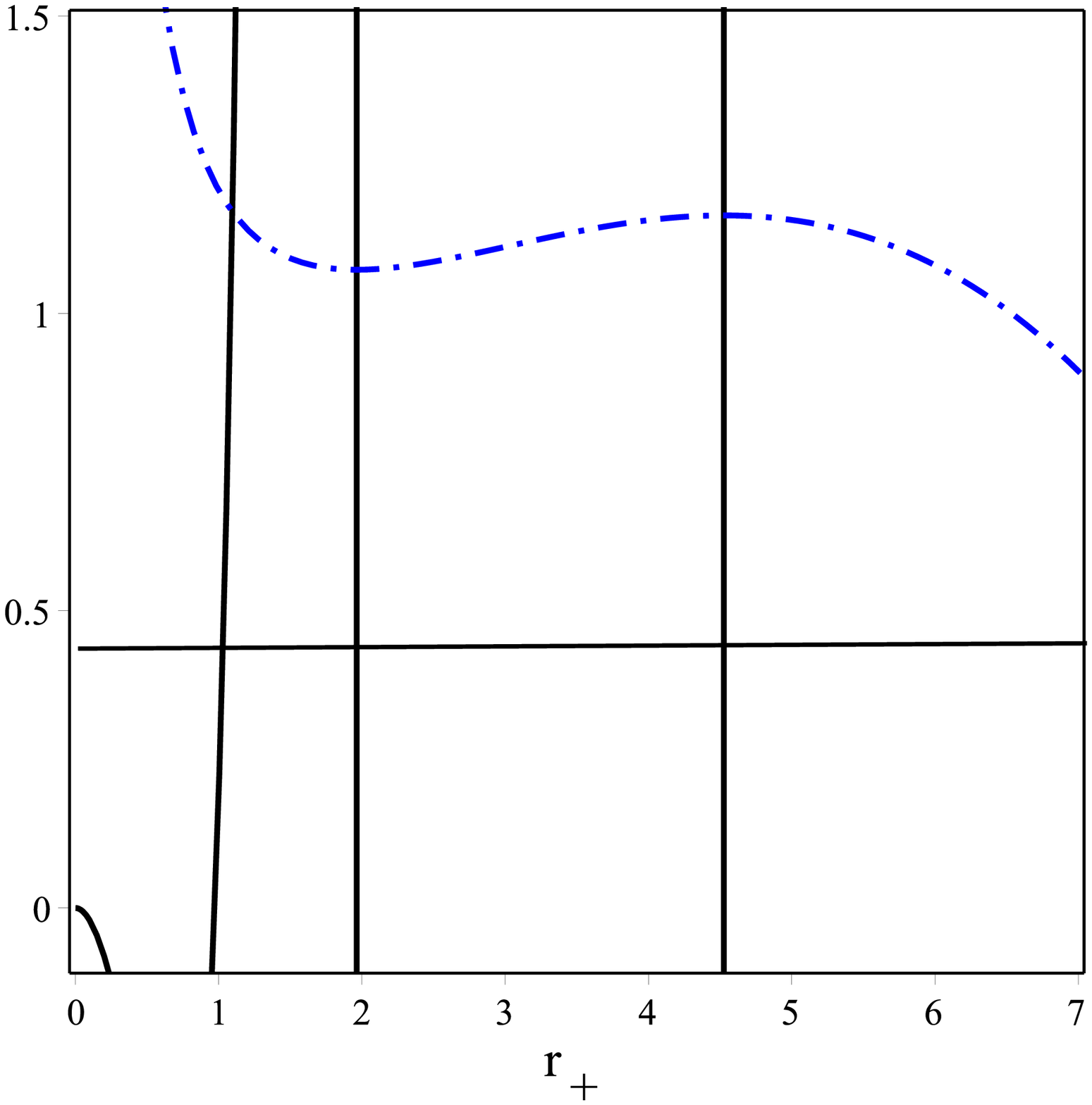}} {\epsfxsize=8cm
\epsffile{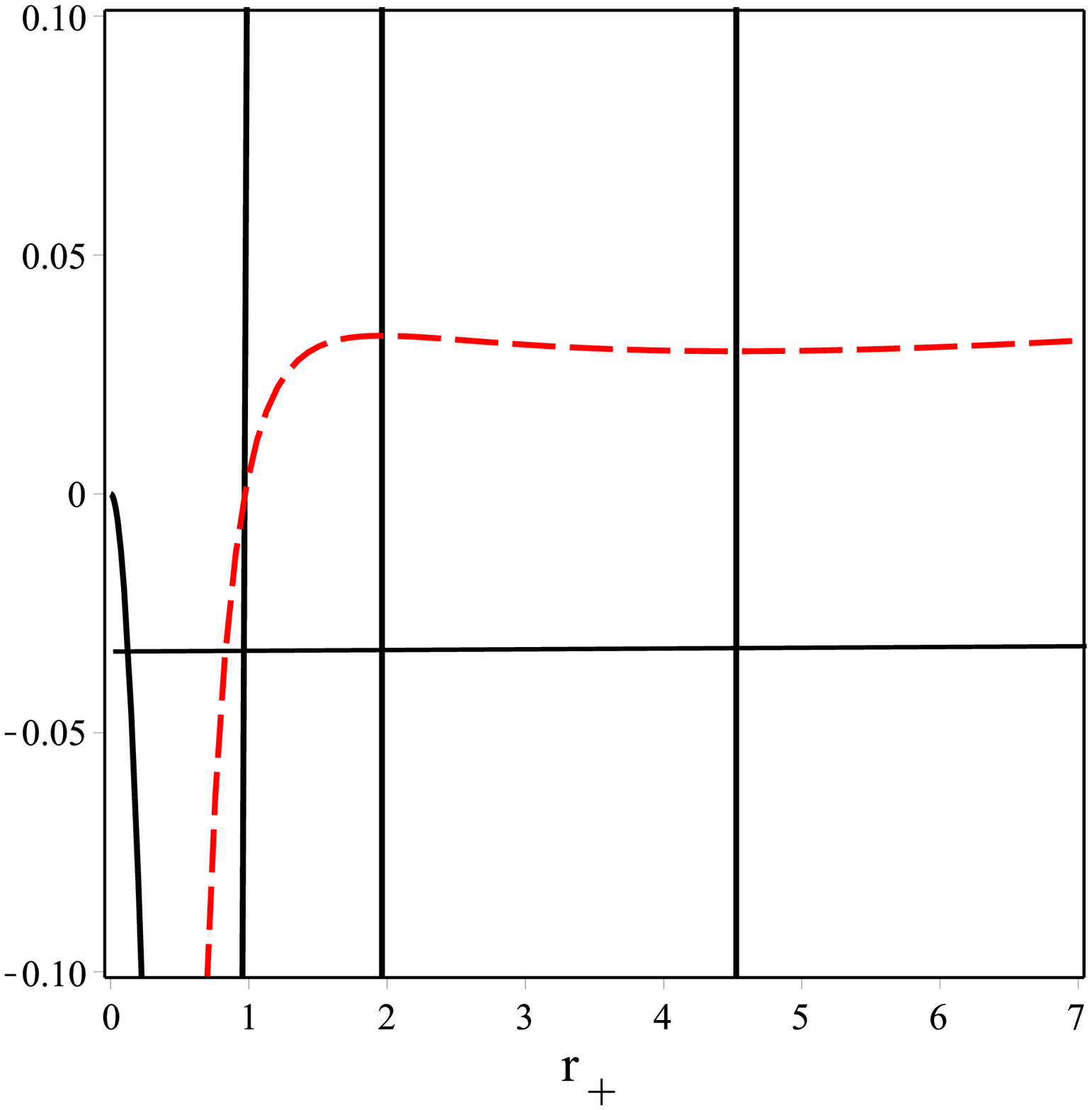}} \caption{The behavior of $
\textcolor{black}{C_{P}}$ ($ \textcolor{black}{solid}$ line), $
\textcolor{red}{T}$ ($ \textcolor{red}{dashed}$ line) and $
\textcolor{blue}{G} $ ($
\textcolor{blue}{dotted\hspace{.1cm}dashed}$ line) in terms of $
r_{+} $ for  $ P=0.0015, Q=1 $.} \label{figcpp}
\end{figure}

\begin{figure}[H]
\centering {\epsfxsize=8cm \epsffile{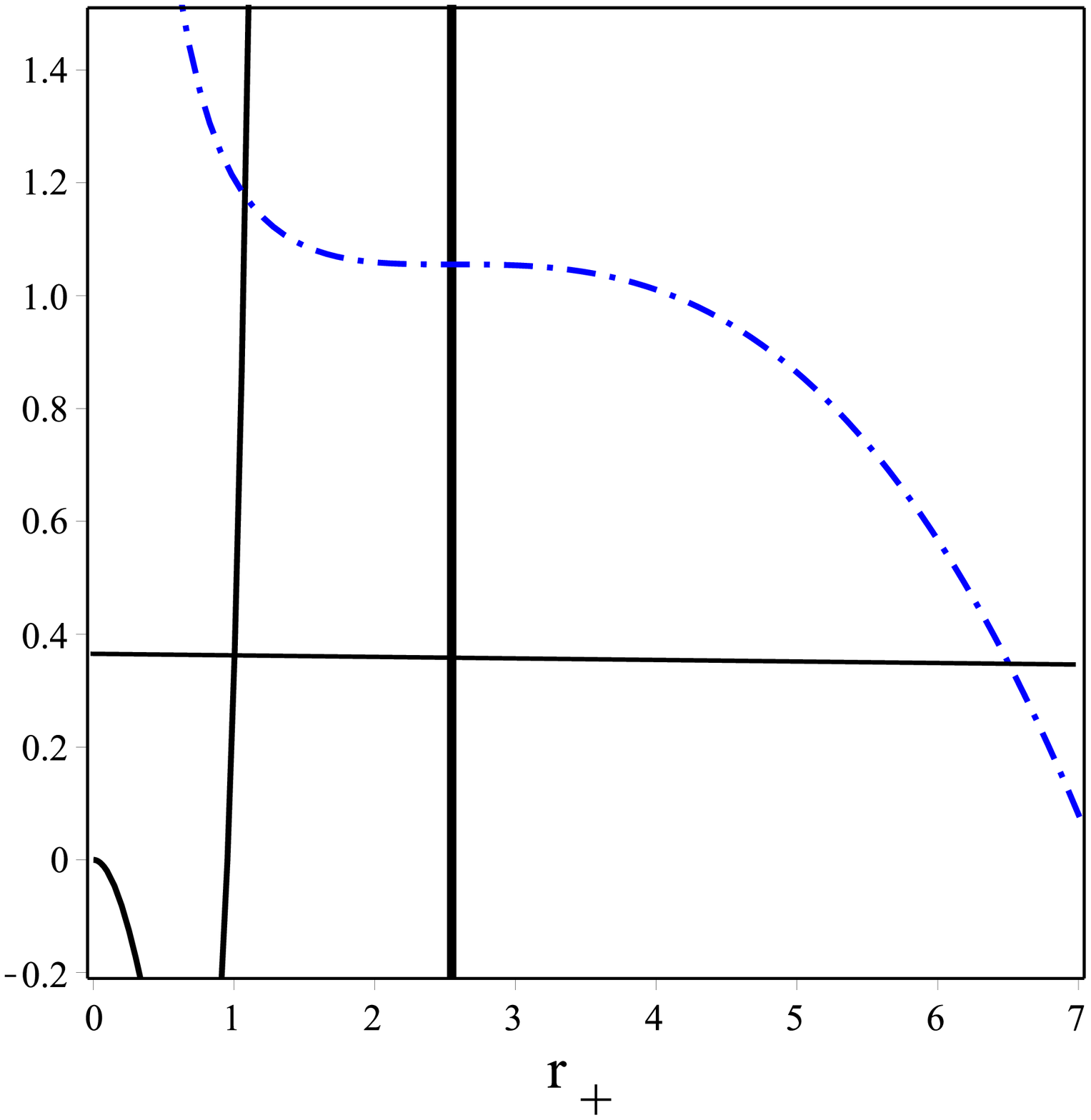}} {\epsfxsize=8cm
\epsffile{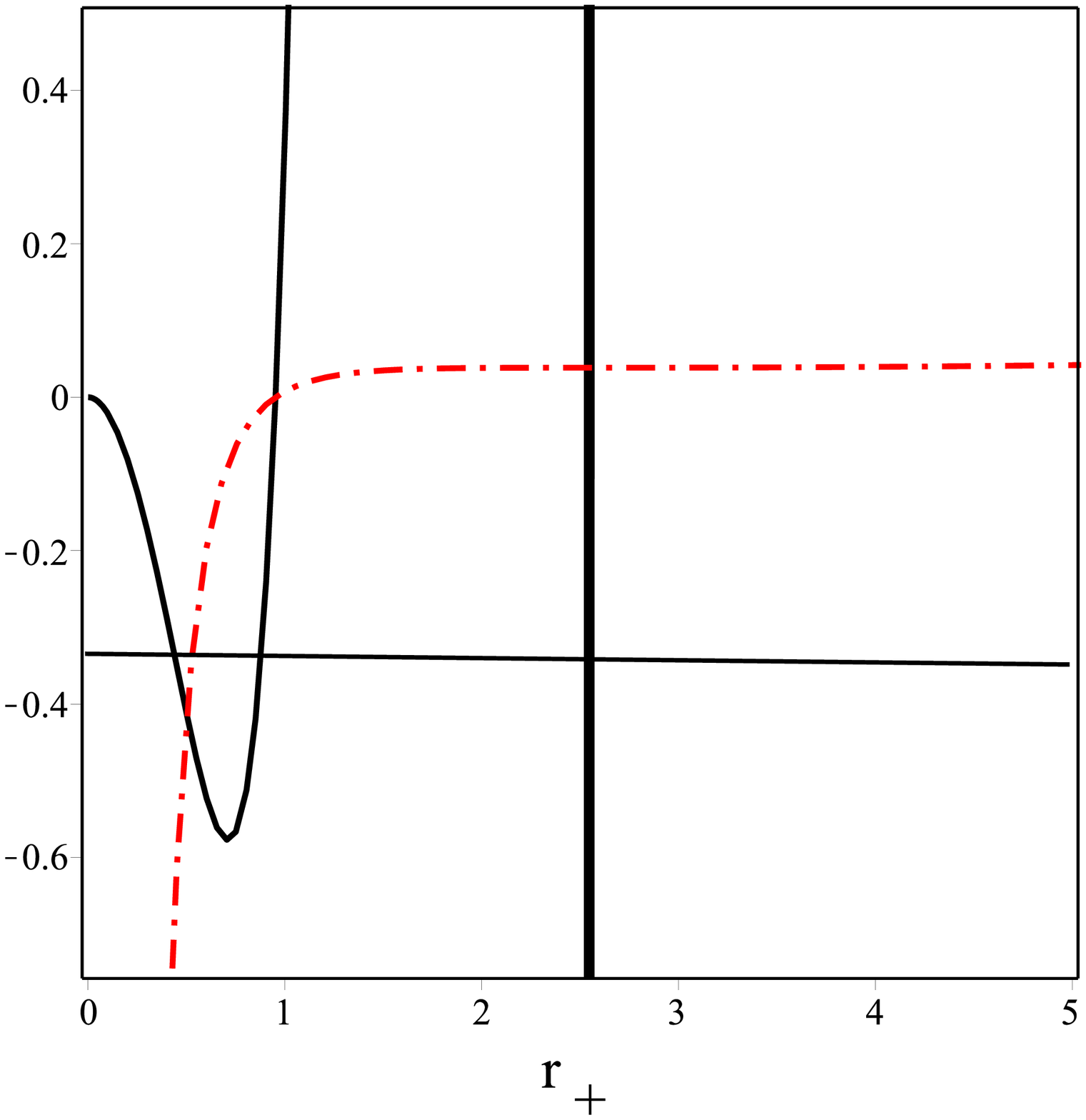}} \caption{The behavior of $
\textcolor{black}{C_{P}}$ ($ \textcolor{black}{solid}$ line), $
\textcolor{red}{T}$ ($ \textcolor{red}{dashed}$ line) and $
\textcolor{blue}{G} $ ($
\textcolor{blue}{dotted\hspace{.1cm}dashed}$ line) in terms of $
r_{+} $ for  $ P_{c}=0.002669, Q=1 $.} \label{figcpp1}
\end{figure}

The behavior of heat capacity could also represent phase transitions \cite{shh, shh1}. By numerical calculation, we have shown the existence of phase transition in Figs. \ref{fig3} and \ref{figcpp}. It is clear that, the heat capacity has two divergencies which separate small stable BHs, unstable region, and large stable ones, and coincidence with extremum points of temperature and Gibbs free energy. Also, the roots of temperature coincide with the points that the heat capacity changes its signature. The roots separate the non-physical solutions with negative temperature from physical black holes with positive temperature. As the pressure approaches the critical pressure, the two divergent points get close to each other. At $ P_{c} $ the divergent points coincide and the intermediate radii black hole would disappear (Fig. \ref{figcpp1}).
\begin{center}
\begin{figure}[H]\epsfxsize=8cm \epsffile{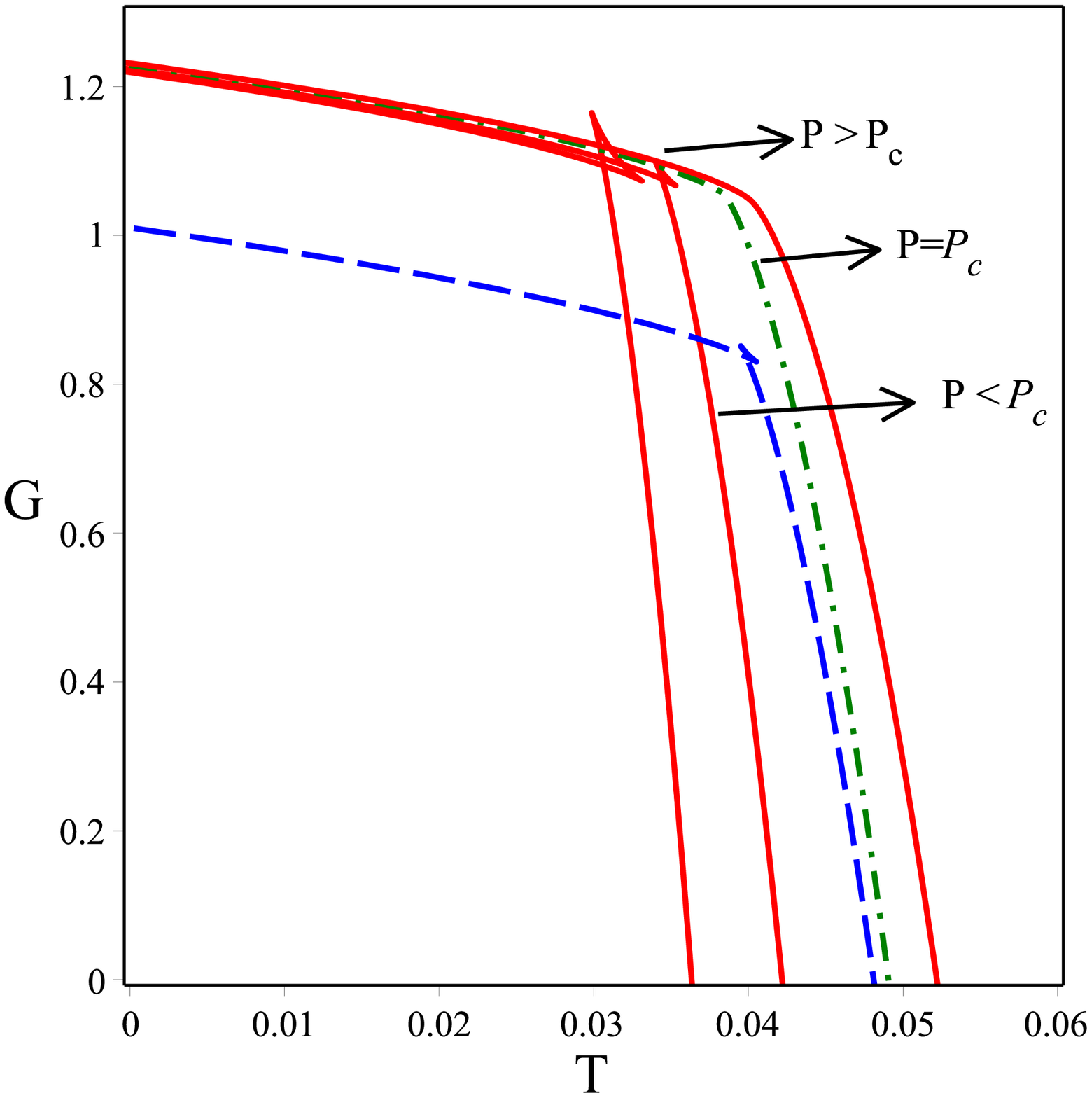}\vspace{0.1cm}\caption{\label{figgt}\small The behavior of $ G $ for our solution in terms of $ T $ for $ P=\textcolor{red}{0.003}, \textcolor{green}{0.002669}, \textcolor{red}{0.002}, \textcolor{red}{0.0015} $ (right to left) and $ Q=1 $, compared to the RN-AdS metric ($ \textcolor{blue}{dashed}$ line) for $  P=\textcolor{blue}{0.002669}$, $ Q=1 $.}
\end{figure}
\end{center}
Gibbs free energy as a function of temperature for various
pressures is shown in Fig. \ref{figgt}. When $
P<\textcolor{green}{0.002669} $, the Gibbs free energy with
respect to temperature develops a swallow-tail like shape. There
is a small/large first order phase transition in the black hole,
which resembles the liquid/gas change of phase occurring in the
Van der Waals fluid. At the critical pressure $
P=\textcolor{green}{0.002669} $, the swallow-tail disappears which
corresponds to the critical point.
\begin{center}
\begin{figure}[H] \epsfxsize=8cm \epsffile{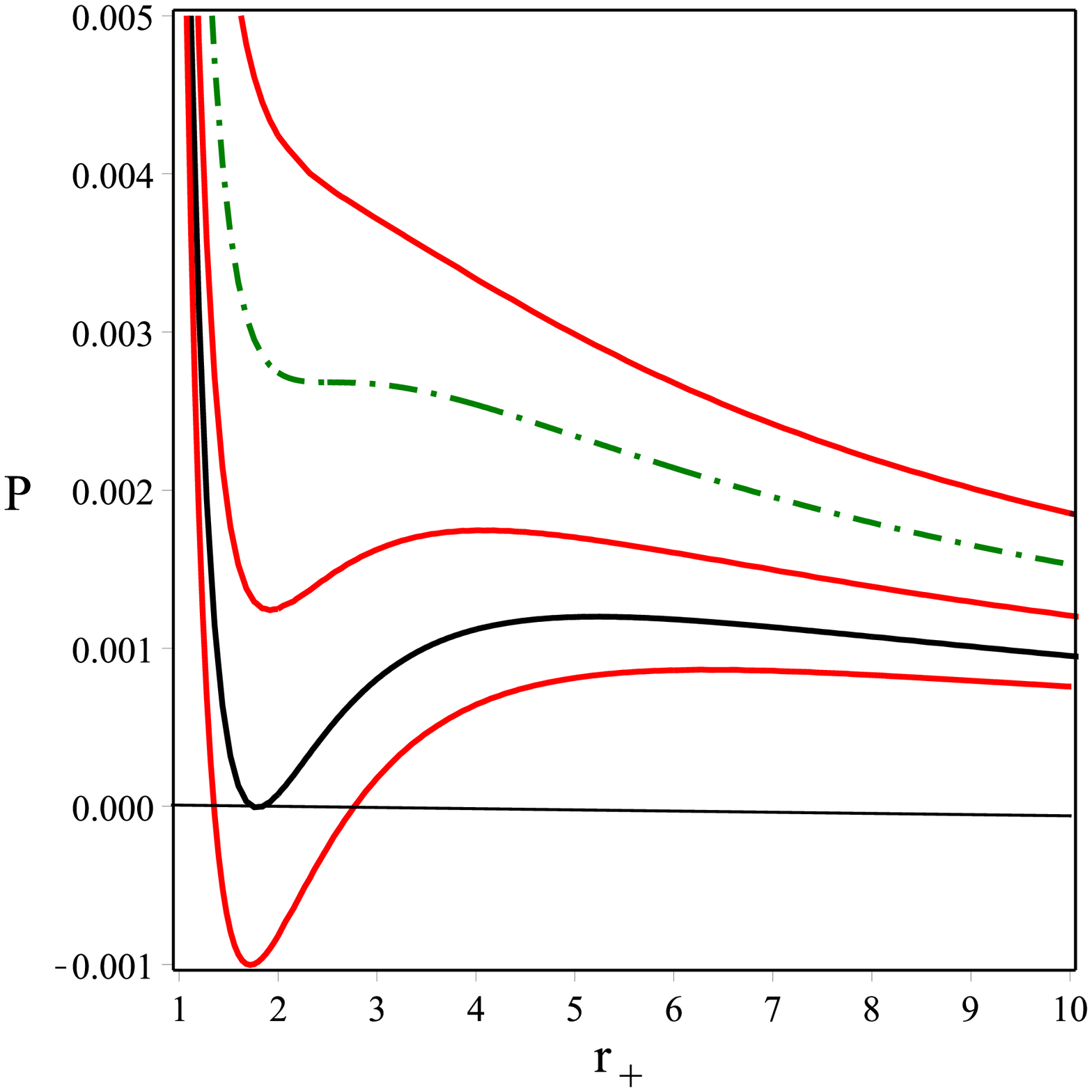}\vspace{0.1cm}\caption{\label{figpv} \small
The behavior of $ P $ in terms of $ r_{+} $ for $ Q=1,
T=\textcolor{red}{0.045}, \textcolor{green}{0.0385},
\textcolor{red}{0.034}, \textcolor{black}{0.0269},
\textcolor{red}{0.023}$ (downwards).}
\end{figure}
\end{center}
In Fig. \ref{figpv}, by numerical calculation, we plot $ P-r_{+}
$, keeping $ T$ and $Q $ fixed. The temperature of isotherm
diagrams decreases from top to bottom. The two upper $
\textcolor{red}{solid} $ lines correspond to the ideal gas phase,
the critical isotherm is denoted by the
$\textcolor{green}{dotted-dashed}$ line, lower solid lines
correspond to temperatures smaller than the critical temperature
and there is also a temperature below which the pressure becomes
negative in a range of $ r_{+} $ (black dashed line). Again, the
resemblance with the liquid/gas phase transition in Van der Waals
gas and also for the RN-AdS black hole is clearly seen. The
critical point is determined by $ \frac{\partial P}{\partial
r_{+}}=0 $ and $ \frac{\partial^{2}P}{\partial r_{+}^{2}}=0 $. By
using eq. (\ref{eq15}), one can find $
\frac{P_{c}r_{c}}{T_{c}}=\frac{(0.002669)(2.54)}{0.0385}=0.176 $
which is independent of charge $ Q $ \cite{6,7,14,cn,nad}.
\section{Quasinormal Modes}\label{qnm}
\begin{figure}[H]
\centering {\epsfxsize=8cm \epsffile{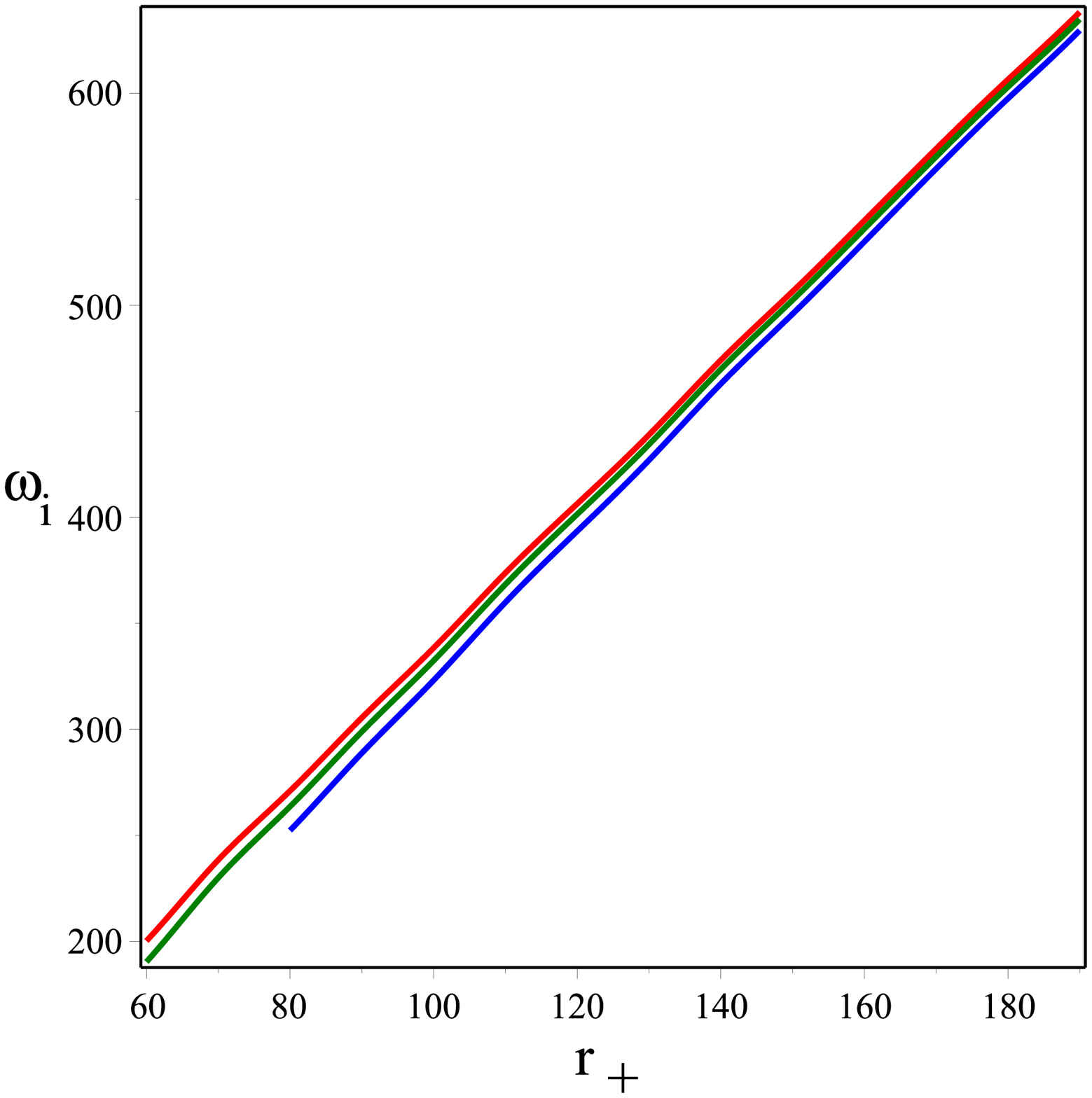}} {\epsfxsize=8cm
\epsffile{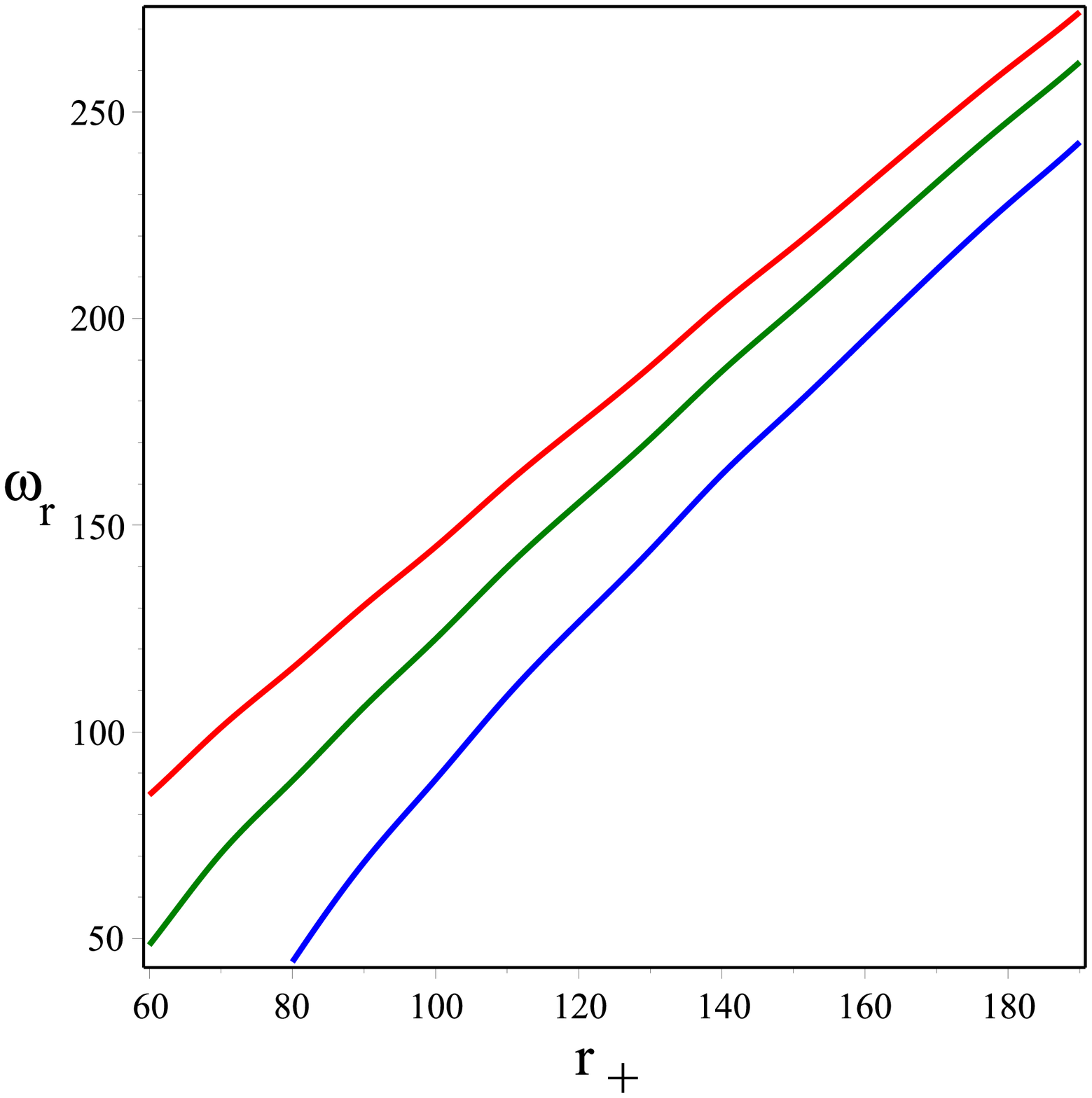}} \caption{The behavior of $ \omega_{i} $ and
$ \omega_{r} $ in terms of $ r_{+} $ for $
Q=\textcolor{red}{5},\textcolor{green}{25},\textcolor{blue}{40},b=1,
l=0$ (downwards)}  \label{fig12}
\end{figure}

In this section, we compute complex frequencies associated with quasinormal modes of the BH by investigating the oscillations of a scalar field. The equation for these perturbations takes the usual form \cite{gth}-\cite{bwe}
\begin{equation}\label{phi}
\frac{1}{\sqrt{\mid g\mid}}\partial _{\mu}(\sqrt{\mid g
\mid}\partial ^{\mu}\phi)=0,
\end{equation}
where g is the determinant of the metric. By decomposing the scalar perturbation and introducing
the tortoise coordinate $ x $ and using spherical harmonics, one can rewrite the radial part of (\ref{phi}) in a Schr${\rm \ddot{o} }$dinger form
\begin{equation}\label{sh}
\left[  -\frac{d^{2}}{dx^{2}}+V(x)-\omega^{2}\right]\varphi(x)=0,
\end{equation}
with $ V $ given by
\begin{equation}
V=f\left(  \frac{l(l+1)}{r^{2}}+\frac{f^{'}}{r}\right),
\end{equation}
and
\begin{equation}
x=\int \frac{dr}{f}.
\end{equation}
The effective potential $ V $ is positive and vanishes at the horizon ($ x=-\infty $). It diverges at $ r=\infty $, which corresponds to a finite value of $ x $,
and hence $ \varphi $ vanishes at infinity. This boundary condition is to be satisfied by the wave
equation of the scalar field. In general, the frequency of waves must be complex, $ \omega=\omega_{r}-i\omega_{i} $. The imaginary and real parts are related to the damping time scale ($\tau_{i}=1/\omega_{i}$) and oscillation time scale ($\tau_{r}=1/\omega_{r}$), respectively. By using the numerical method suggested in \cite{gth}-\cite{bwe}, we have computed the quasinormal frequencies via expanding the solution around the horizon and imposing the boundary condition that the solution vanishes at infinity.
Here, we have considered large black holes, because large black holes in AdS-CFT correspond to a thermal state, and its perturbation corresponds to the perturbation of thermal state, and damping of perturbation in AdS is translated as return to equilibrium in thermal state. Small black hole is unstable and does not correspond to any thermal state in CFT \cite{rak2}.
For large black holes ($ r_{+}\gg b $), the relation of the values of the lowest quasinormal frequencies for $ l=0 $ and selected values of $ r_{+} $ for different charge $ Q $ are exhibited in Fig. \ref{fig12}. It can be seen that
the real and the imaginary parts of the frequency are not linear functions of $ r_{+} $ similar to RN-AdS \cite{bwe}. This can be better seen by calculation of diagram slope from Table \ref{tab:tank}.
From Fig. \ref{fig12}a, we learn that as $ Q $ increases,
$ \omega_{i} $ and $ \omega_{r} $ decrease. According to the AdS/CFT correspondence, this means that for large $Q$, it is slower for the quasinormal ringing to settle
down to thermal equilibrium and also, the frequency of the oscillation becomes smaller (Fig \ref{fig12}b).
\begin{figure}[H]
\centering {\epsfxsize=8cm \epsffile{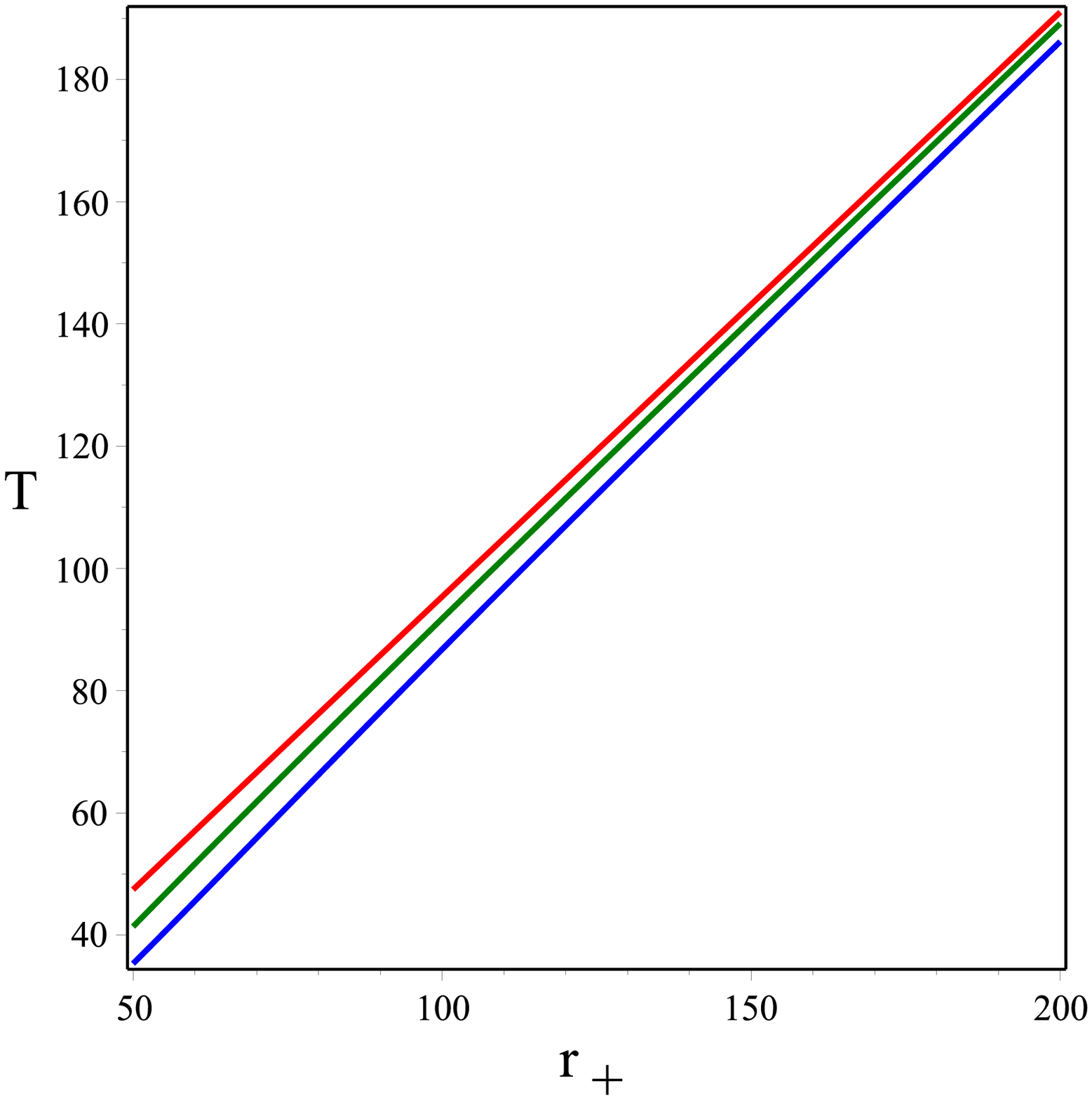}} {\epsfxsize=8cm
\epsffile{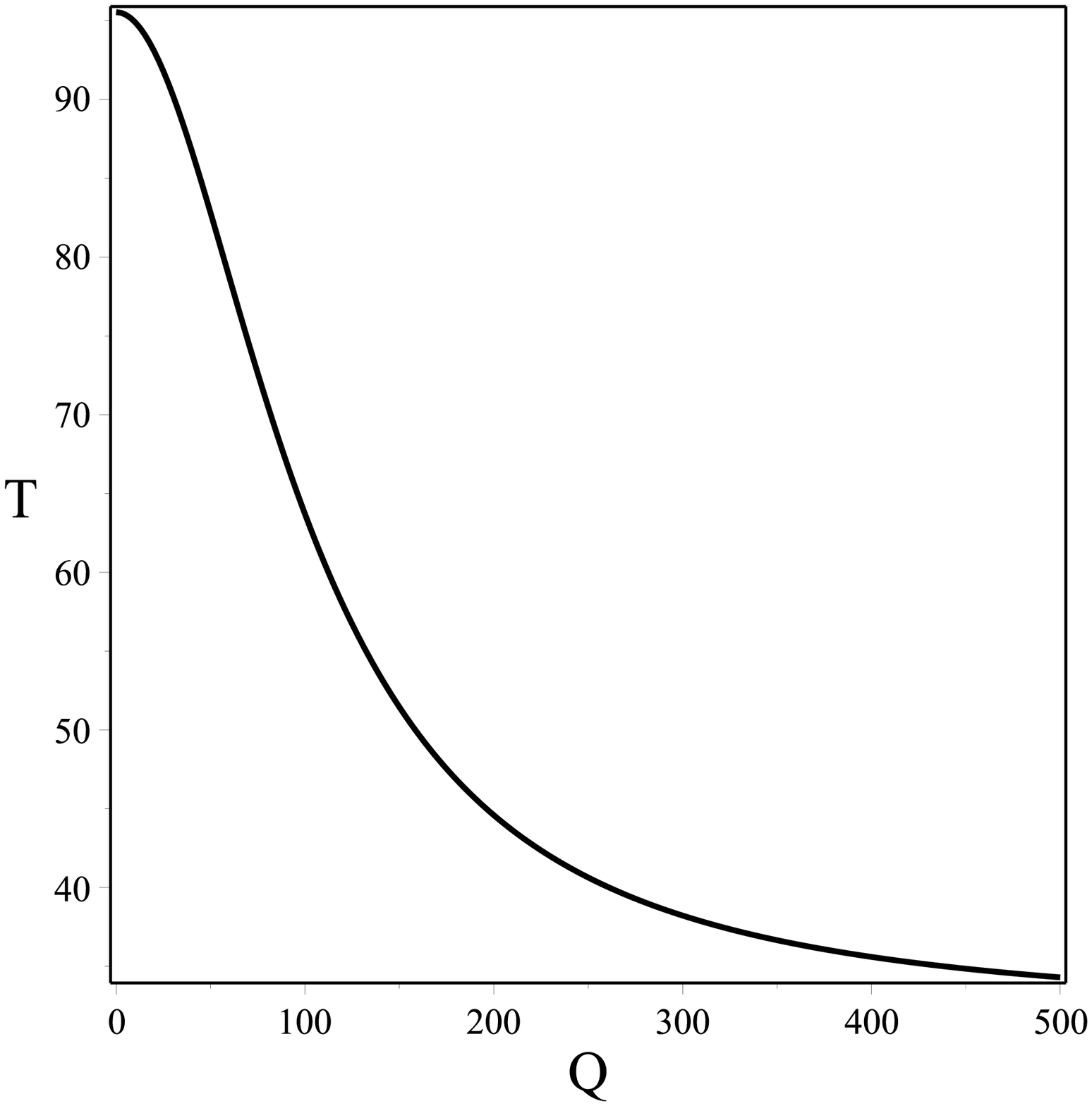}} \caption{The behavior of $ T$ in terms of $
r_{+} $ for
$Q=\textcolor{red}{5},\textcolor{green}{25},\textcolor{blue}{40}$
(left), and the behavior of $T$ in terms of $ Q$ for $b=1, l=0$
(right).}  \label{fig17}
\end{figure}
From eq. (\ref{eq15}) and Fig. \ref{fig17}, one can find that the temperature is not a linear function of the event horizon radius. As $ Q $ increases, the linear relation between $ T $  and $ r_{+} $ no longer holds. Again, the linear relation between real and imaginary parts of frequency with charge is changed (Fig.\ref{fig13}).
\begin{figure}[H]
\centering {\epsfxsize=8cm \epsffile{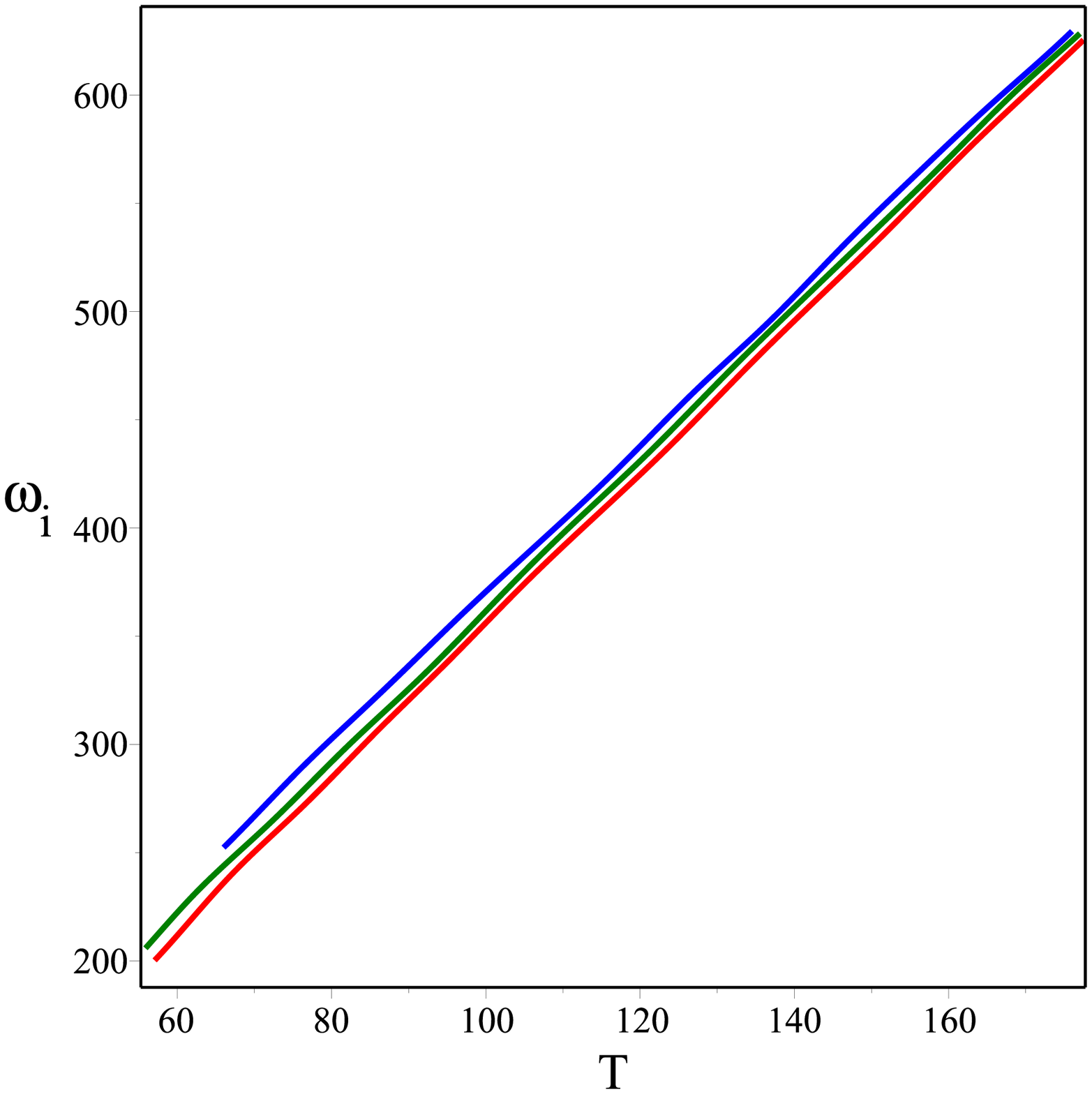}} {\epsfxsize=8cm
\epsffile{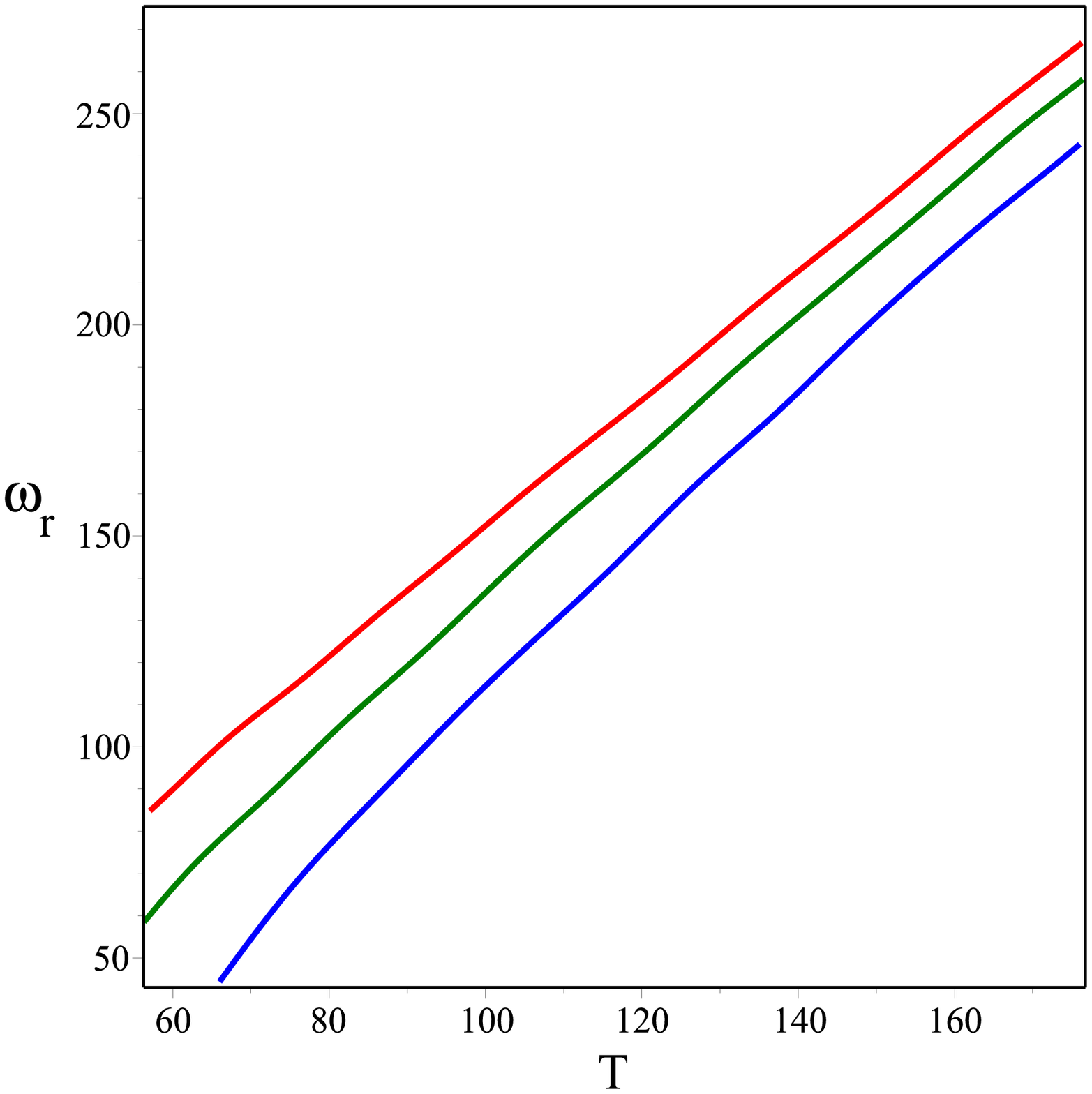}} \caption{The behavior of $ \omega_{i} $ and
$ \omega_{r} $ in terms of $ T $ for $
Q=\textcolor{red}{5},\textcolor{green}{25},\textcolor{blue}{40},b=1,
l=0$.}  \label{fig13}
\end{figure}
In Fig. \ref{fig14}, the dependence of $ \omega_{r} $ and $ \omega_{i} $ on $ l $ for different values of $ Q $ is exhibited. Unlike the case of RN-AdS \cite{bwe} and Schwarzschild-AdS \cite{gth} black holes, by increasing $l$, $ \omega_{i} $ and  $ \omega_{r} $ increase and different values of $ Q $ do not change the qualitative
behavior of $ \omega_{r} $ and $ \omega_{i} $ with $ l $.
\begin{figure}[H]
\centering {\epsfxsize=8cm \epsffile{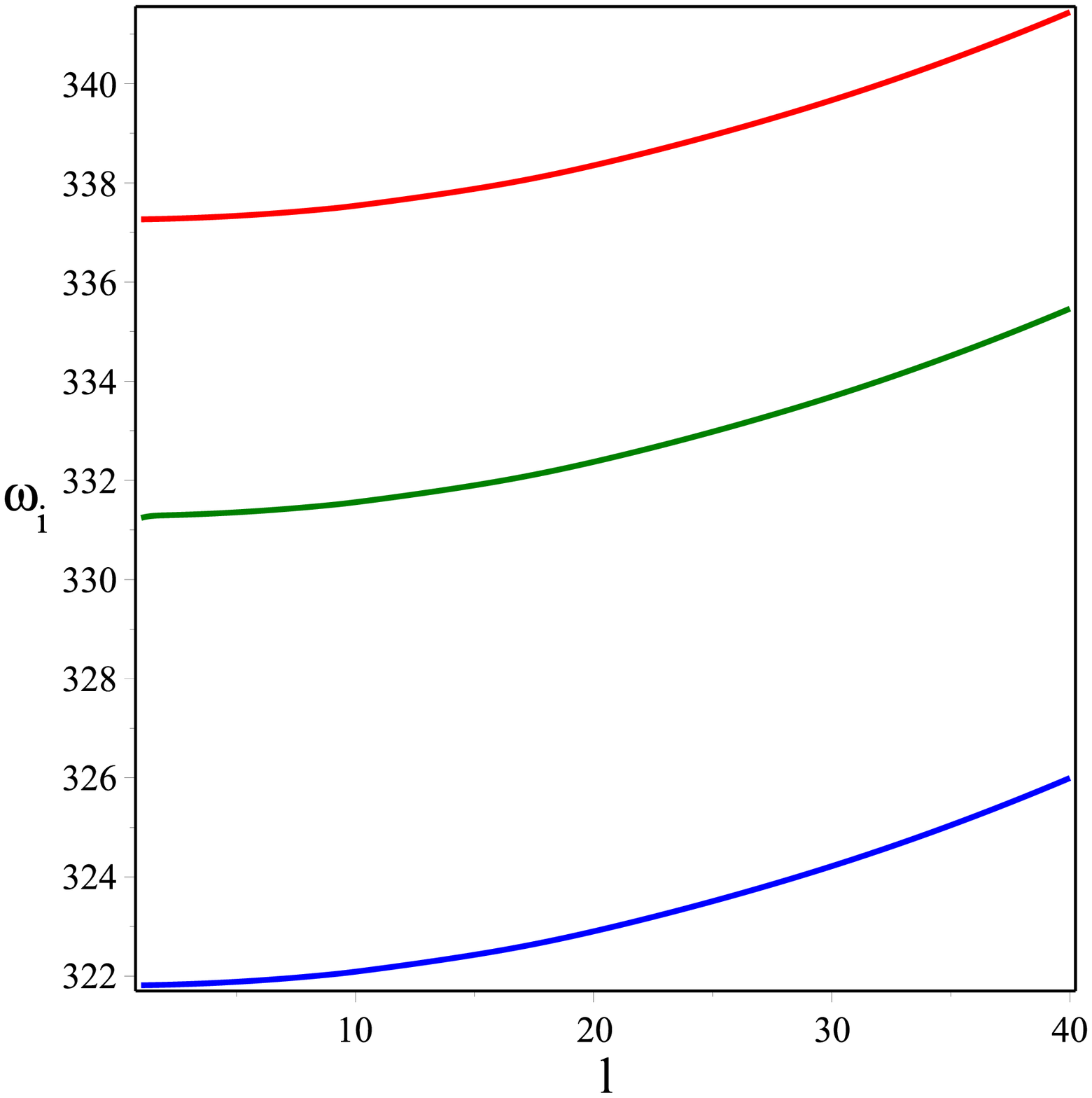}} {\epsfxsize=8cm
\epsffile{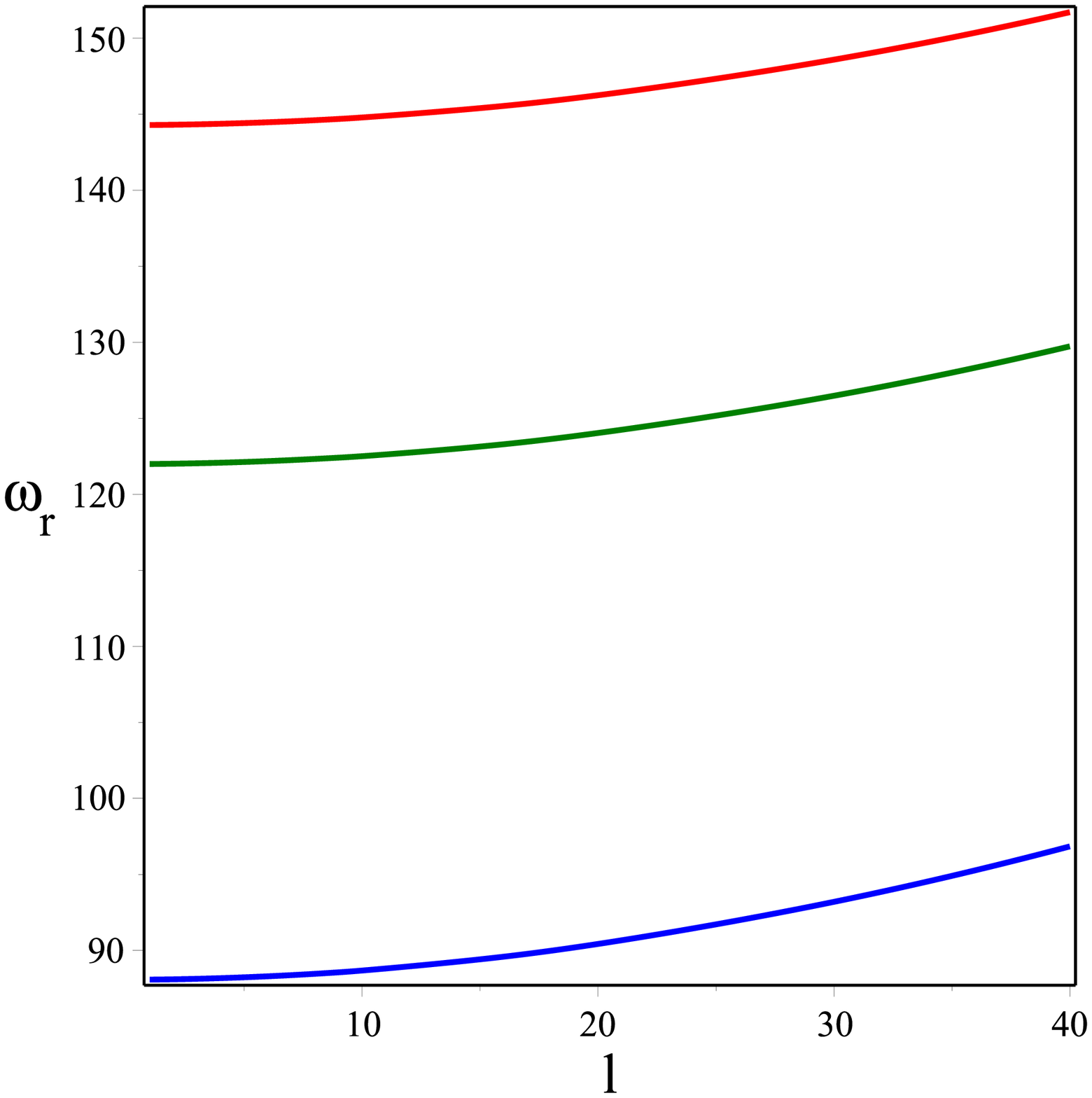}} \caption{The behavior of $ \omega_{i} $ and
and $ \omega_{r} $ in terms of $ l $ for $ Q=\textcolor{red}{5},
\textcolor{green}{25},\textcolor{blue}{40}, b=1, r_{+}=100$
(downwards).} \label{fig14}
\end{figure}
From Fig. \ref{fig15}, as the black hole charge $Q$ increases, the real and imaginary parts of the large black hole QNM frequencies decrease. Also, for fixed charge, as $ l $ increases, QNM frequency increases, and this is consistent with Fig. \ref{fig14}.
\begin{figure}[H]
\centering {\epsfxsize=8cm \epsffile{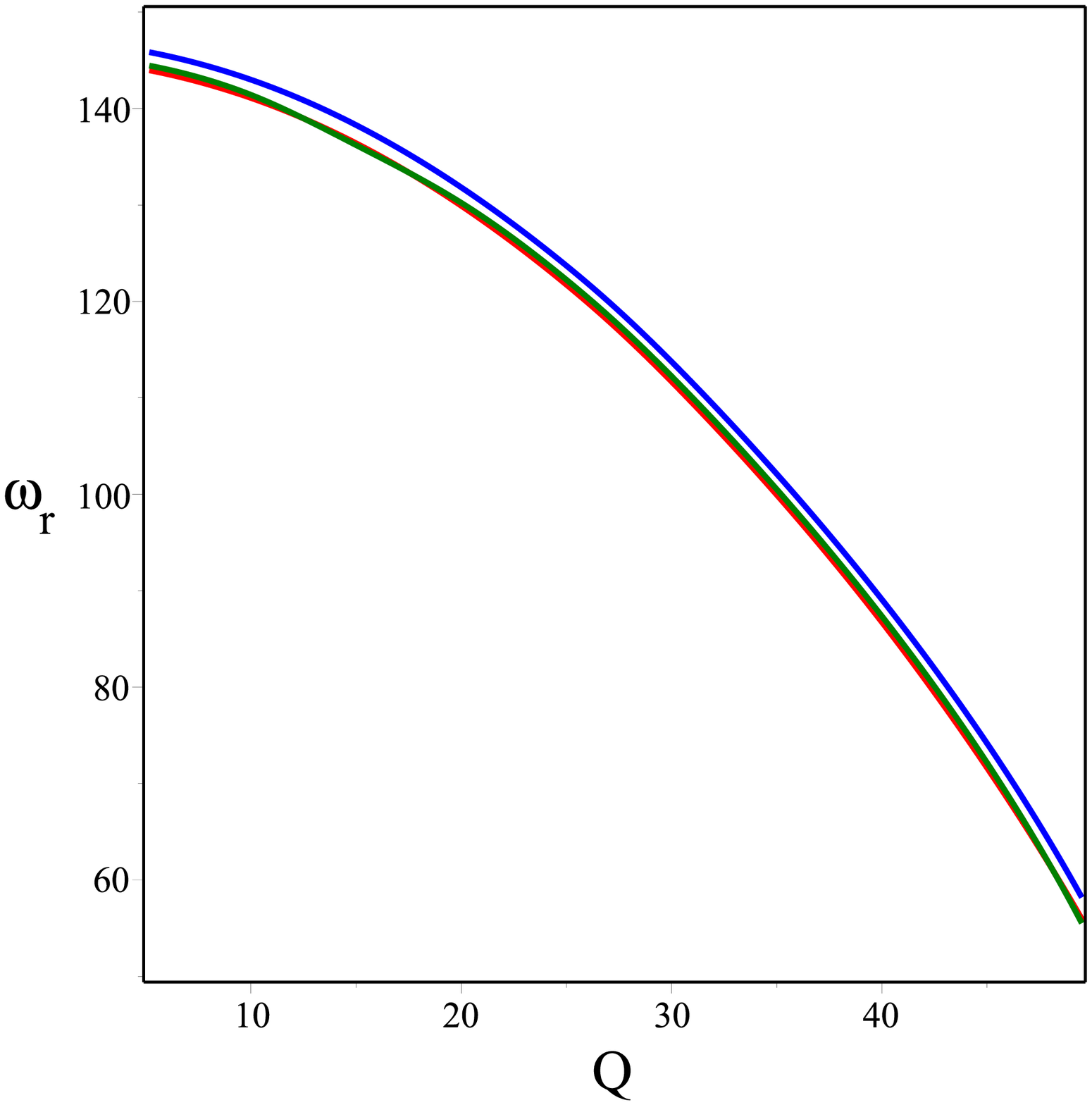}} {\epsfxsize=8cm
\epsffile{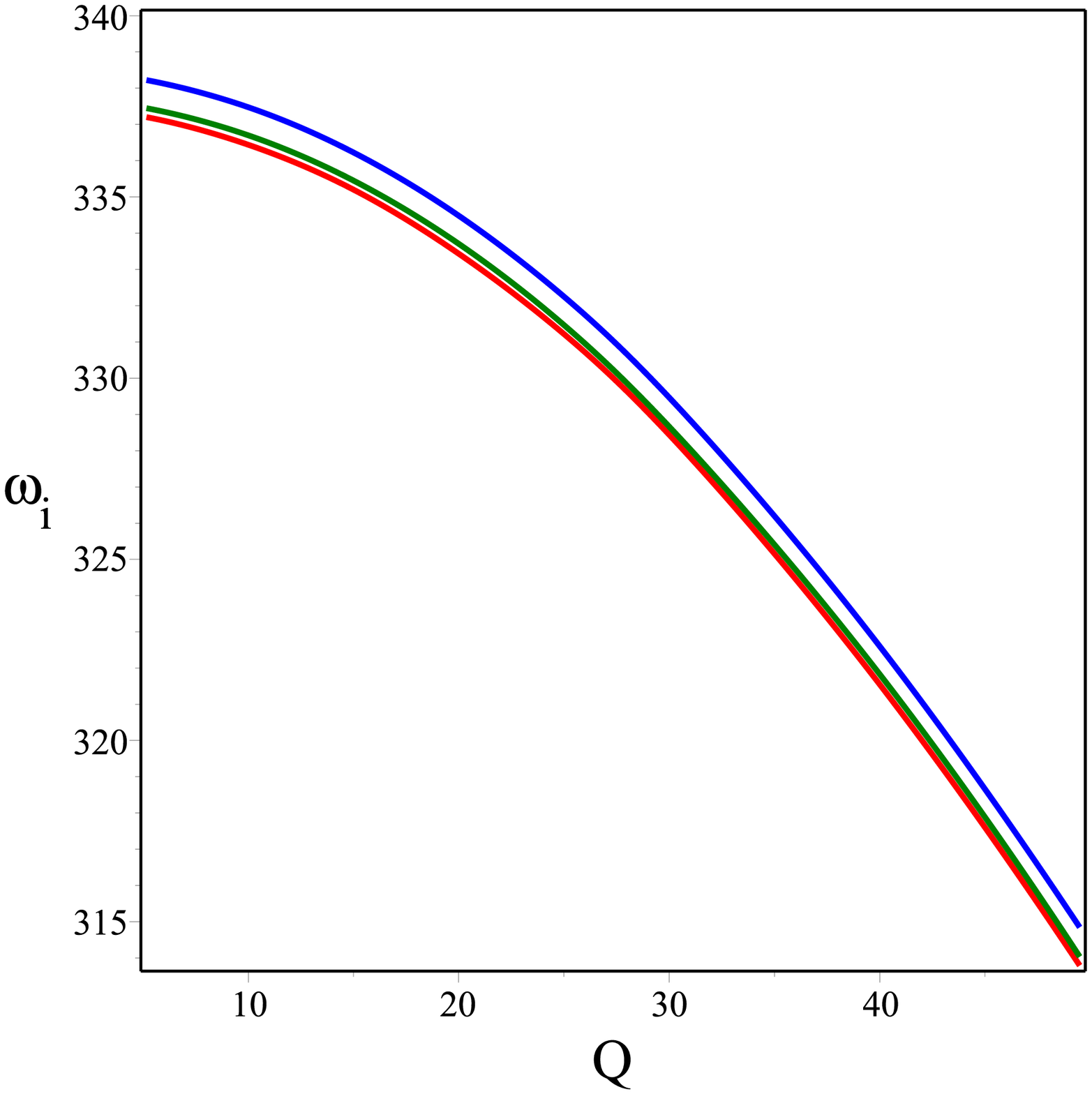}} \caption{The behavior of $ \omega_{r} $ and
$ \omega_{i} $ in terms of $ Q $ for $
l=\textcolor{red}{0},\textcolor{green}{10},\textcolor{blue}{20},
b=1, r_{+}=100$, (upwards).} \label{fig15}
\end{figure}
\begin{table}[H]
\begin{center}
\setlength{\tabcolsep}{1cm}
\begin{tabular} { c   c  c  c   }
\hline
$r_{+}$ &Q=5 &Q=25 &   Q=40   \\ \hline
 50 & 70.71+168.3 i &27.93+156.8 i& \\

60 & 84.00+198.1 i  &47.64+18.2 i&  \\

70&102.40+240.7 i&72.10+232.4 i&17.91+220.0 i \\

80 &115.00+269.7 i&87.60+262.3 i& 44.70+ 250.8 i\\

90 &131.00+306.5 i&106.60+299.0 i& 69.50+289.7 i\\

100 &144.30+337.3 i&122.0+331.3 i& 88.00+321.8 i\\

150 &217.10+505.8 i&201.90+501.8 i&178.20+495.4 i \\

200 & 290.00+674.9 i&278.50+679.1 i&260.00+667.0 i \\
\hline
\end{tabular}
\end{center}
\caption{Quasinormal frequencies of the scalar perturbations for $l=0 $ and $ b=1 $. \label{tab:tank}}
\end{table}
\begin{figure}[H]
\centering
 {\epsfxsize=8cm \epsffile{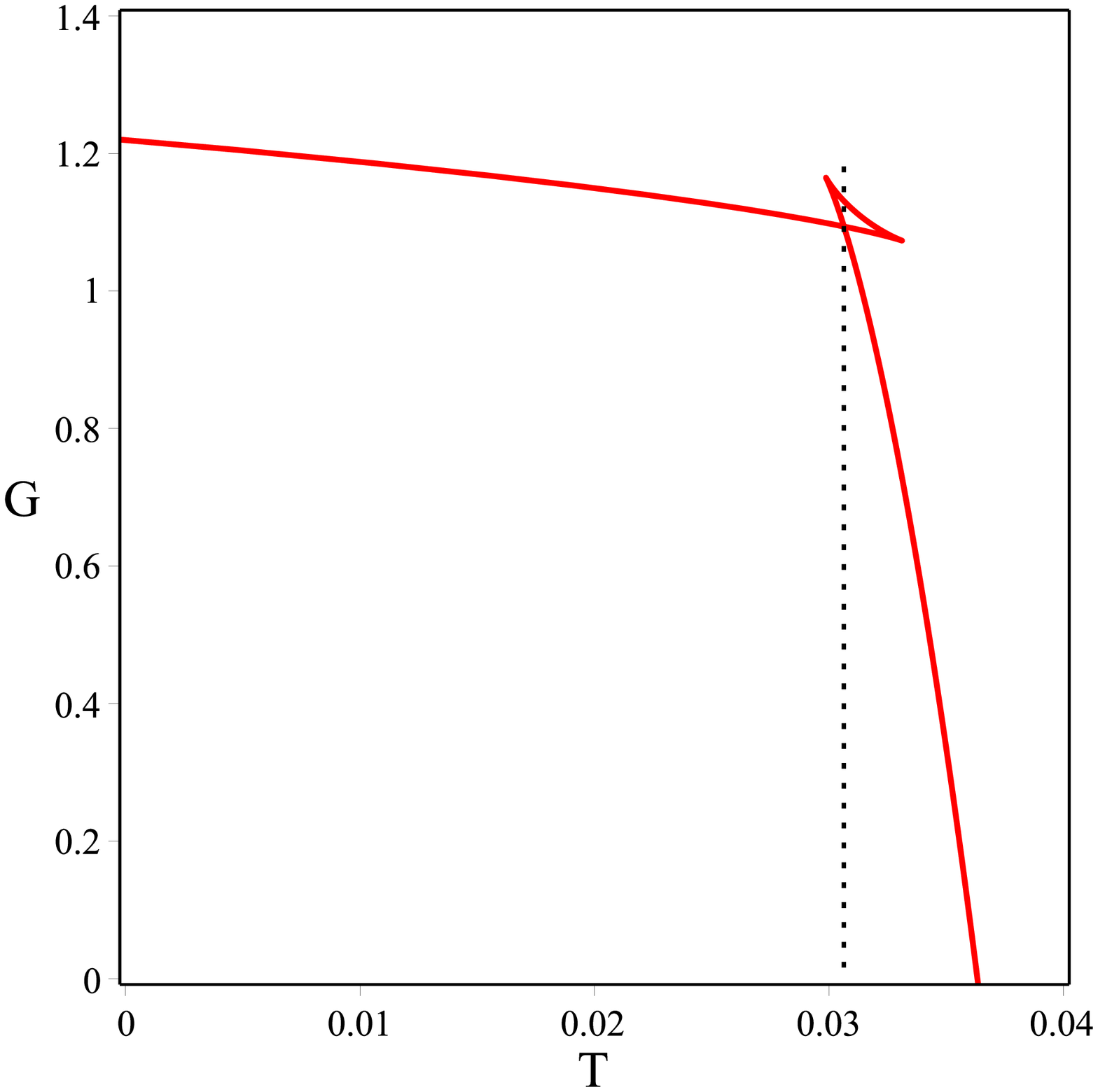}}
 {\epsfxsize=8cm \epsffile{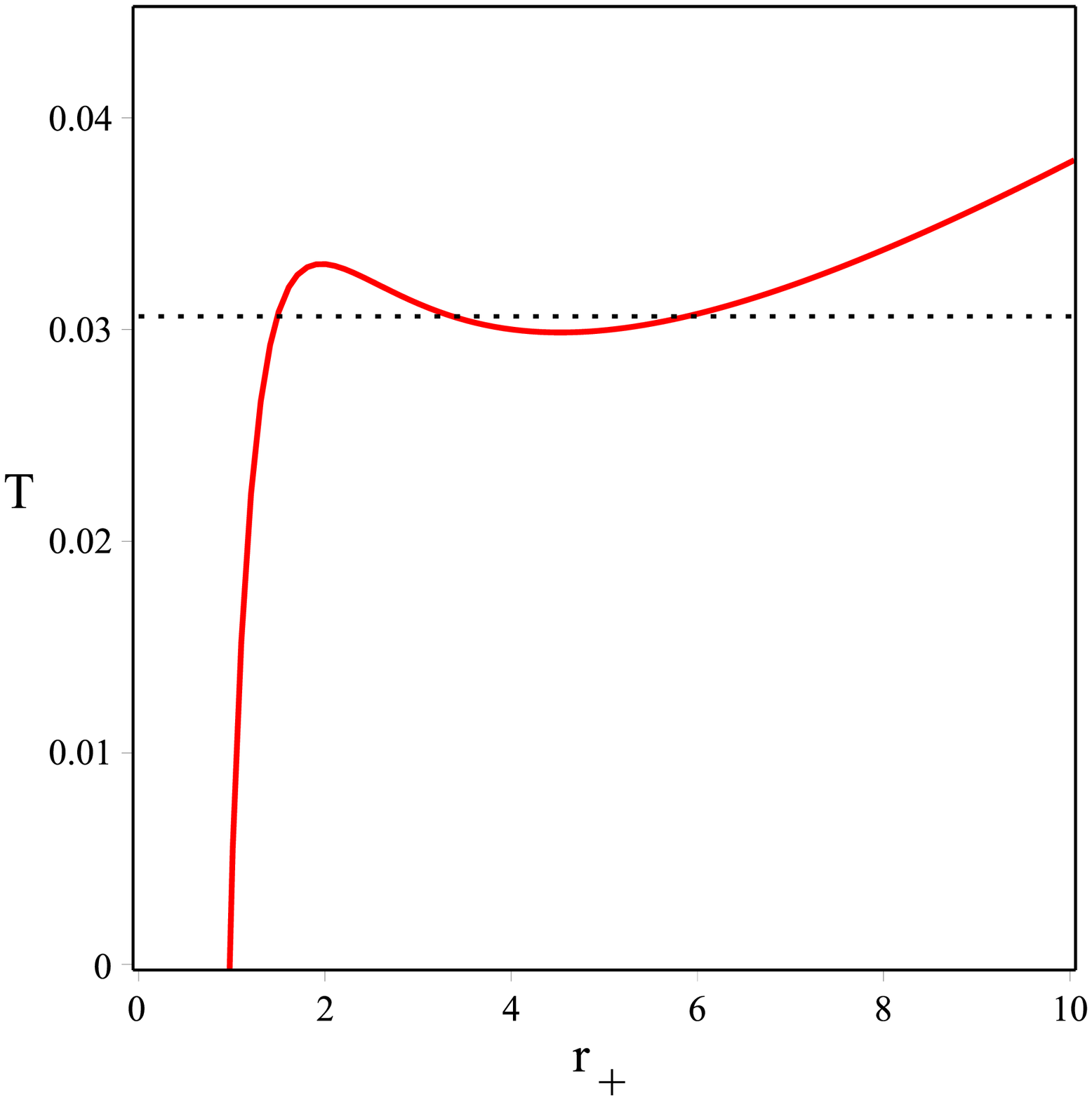}}
\caption{The Gibbs free energy is given as a function of
temperature for $Q=1, P =0.0015$ and $T_{c}=0.0306$ (left). The
behavior of black hole temperature $T$ as a function of the black
hole horizon $r_{+}$ for $Q=1, P =0.0015$ (right).} \label{fig16}
\end{figure}
\begin{figure}[H]
\centering
 {\epsfxsize=8cm \epsffile{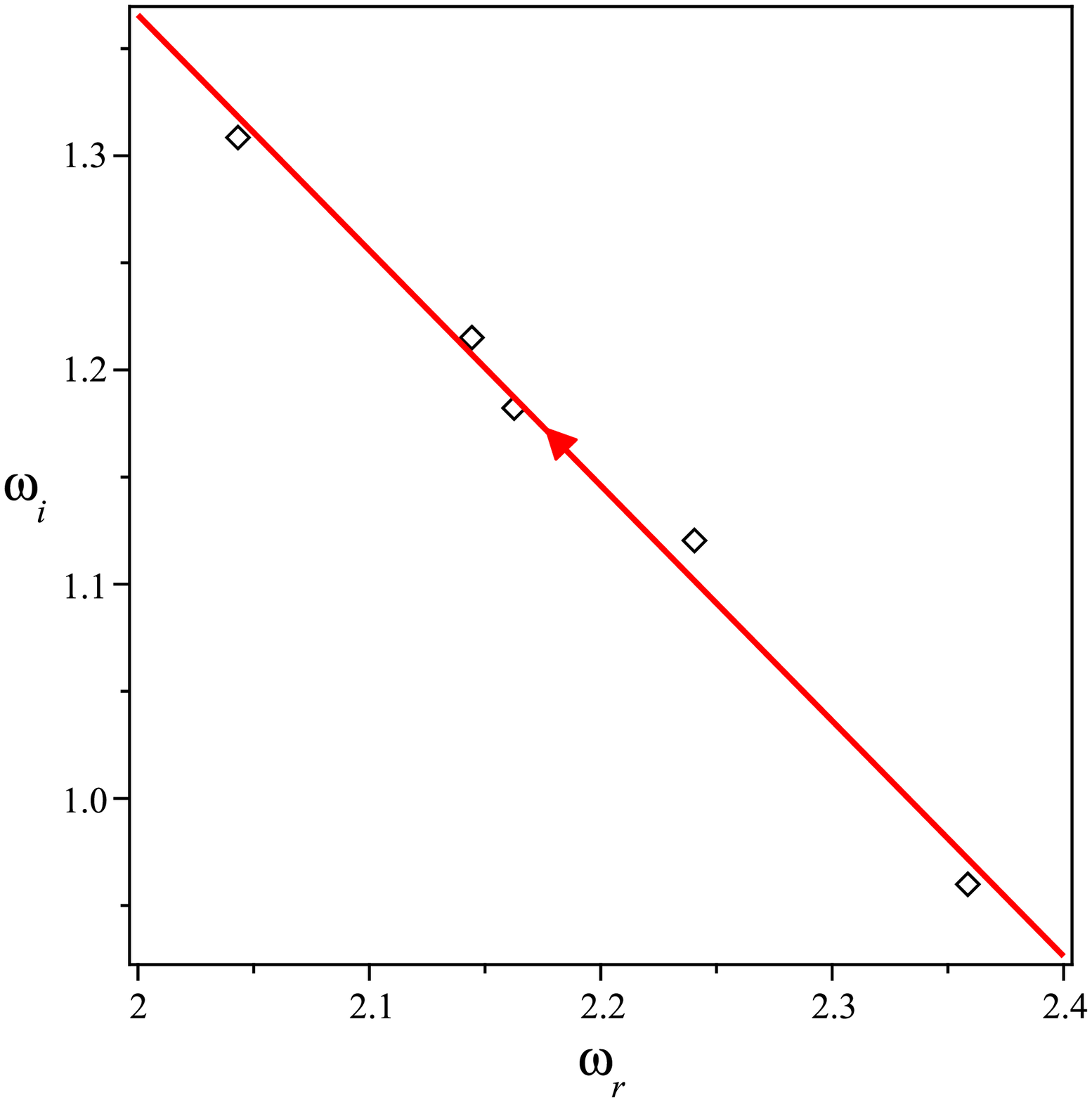}}
 {\epsfxsize=8cm \epsffile{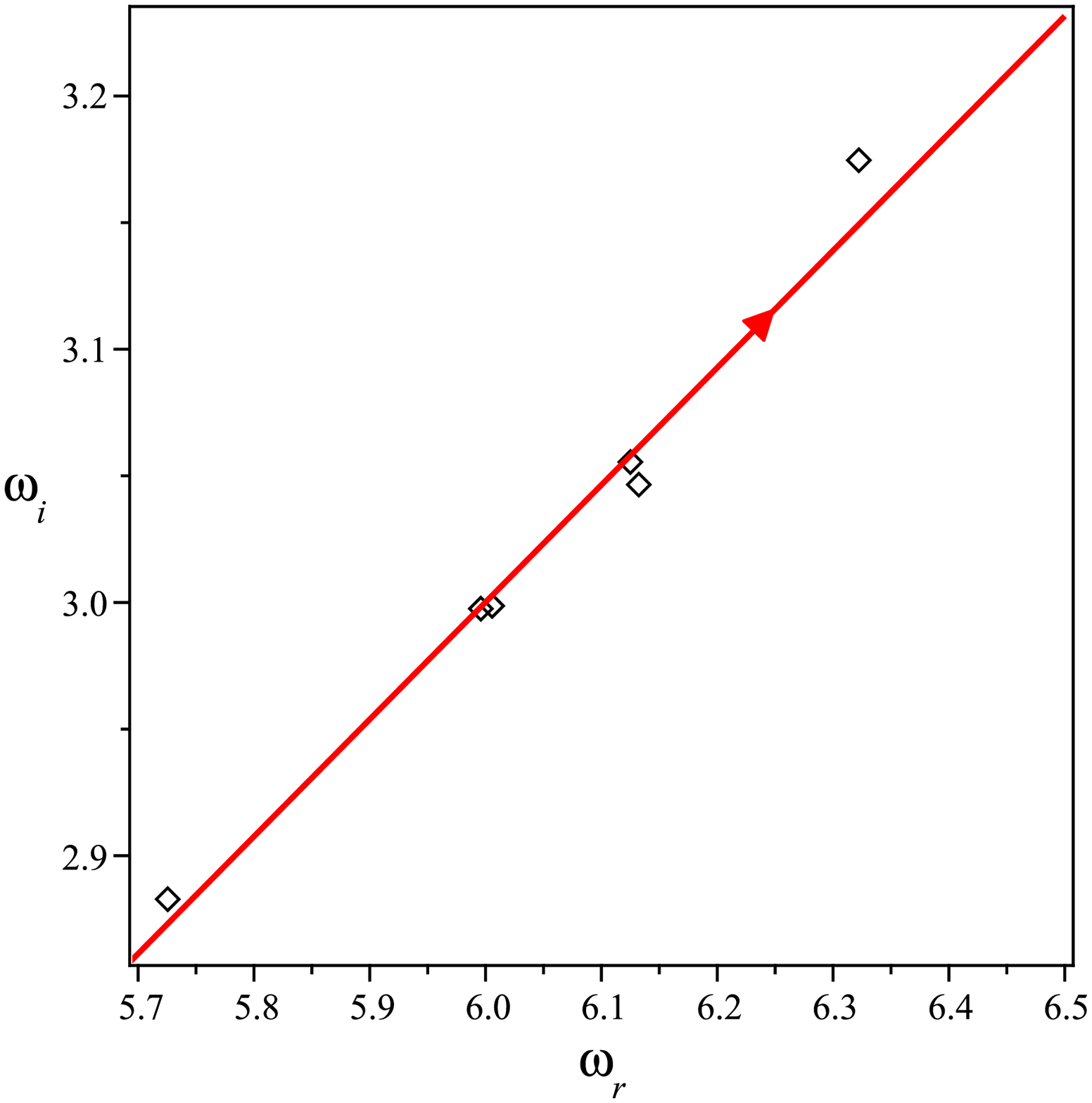}}
\caption{ The behavior of quasinormal modes for small black holes
(left) and for large black holes (right). The arrow indicates the
increase of black hole horizon.} \label{fig17}
\end{figure}

Finally, let us investigate whether the signature of
thermodynamical first order phase transition in charged AdS black
holes can be reflected in the dynamical quasi normal modes
behavior in the massless scalar perturbation. We study the
dynamical perturbations in isobaric process by fixing the pressure
$P$ to obtain the phase transitions between small-large black
holes. In Fig. \ref{fig16}, we have plotted the behavior of an
isobar. The crossing curves in the left plot indicates the
coexistence of two phases in equilibrium that correspond to the
separated points in the right plot for $T_{c}=0.0306$. The quasi
normal frequencies corresponding to small-large black holes have
been plotted in Fig. \ref{fig17}. For the case of small black
holes, one can see that by increasing the radius of event horizon
the imaginary part increases and real part of frequencies
decreases. While, for the large black hole, we find that when the
black hole radius increases, the real part together with the
imaginary part of quasinormal frequencies increase
\cite{Konoplya:2017zwo},\cite{Prasia:2016fcc}.
\begin{table}[H]
\begin{center}
\setlength{\tabcolsep}{1cm}
\begin{tabular} { c   c  c }
\hline
 T&$r_{+}$ &$\omega$  \\ \hline
 0.0265 & 1.30  &2.3562+0.9616 i \\

0.0280 & 1.35 &2.2392+1.1242 i \\

0.0290&1.40&2.1612+1.1845 i \\

0.0299&1.45&2.1420+1.2171 i\\

0.0305&1.49&2.0415+1.3105 i\\
\hline
0.0307 &6.00&5.7229+2.8845 i\\

0.0310&6.50&5.9928+2.9997 i \\

0.0330 & 7.50&6.0015+3.0003 i \\

0.0340 & 8.00&6.1204+3.0572 i \\

0.0350 & 8.50&6.1297+3.0149 i \\

0.0360 & 9.00&6.3184+3.1765 i \\
\hline
\end{tabular}
\end{center}
\caption{ The quasinormal frequencies of massless scalar perturbation. The upper part, above the horizontal line, is for the small black hole phase and the lower part is for the large black hole phase
 for $l=0 $ and $P=0.0015$. \label{tab:tank1}}
\end{table}
\section{Conclusion}\label{sec4}
Although black hole singularities seem to be inevitable within general theory of relativity and through the collapse of ordinary matter which respects energy conditions, there are alternative models which avoid the spacetime singularity by ignoring energy conditions or other assumptions of the singularity theorems \cite{29,30,31}. In the present work, we demonstrated that a non-linear electromagnetic Lagrangian can lead to black hole solutions which behave like a Reissner-Nordestr$\ddot{o}$m-AdS BH at large distance, while having a quasi non-singular core in the sense that the metric function behaves like $ f(r)\approx 1-\frac{2\alpha}{\beta^{2}}r+O(r^{2}) $ as $ r \rightarrow 0$. The electromagnetic Lagrangian, although having a relatively complicated form, reduces to the Maxwellian form as $ F\rightarrow 0 $. The BH solutions presented here have the interesting property that their ADM mass and charge are not free parameter, but depend on the Lagrangian parameters $ \alpha $ and $\beta$.\\
Since, thermodynamical behavior are of great importance in search for a quantum theory of gravitation, we have also managed to perform a thermodynamic investigation of the BH solutions. The conserved and thermodynamic quantities were calculated and the validity of the first law was checked. Global stability of the BH was examined by plotting the Gibbs free energy, and the heat capacities were studied to check the local stability. We showed that the present solutions admitted small/large phase transitions similar to the Van der Waals liquid/gas phase transition. Then, by writing and solving the wave equation for a scalar field in the BH background spacetime, the quasi-normal modes were calculated. We pointed out that the frequencies of quasi normal modes behave somehow differently form those of RN-AdS and Schwarzschild-AdS black holes. The effect of the BH charge on eigen-frequencies were demonstrated through several diagrams which were calculated numerically.\\  Finally, we have obtained the quasinormal frequencies of massless scalar perturbations around small and large black hole. We found that when the Van der Waals-like phase transition happens, as the horizon radius increase, the slopes of the quasinormal frequency change differently in the small and large black holes.\\ As a further remark, in order to more study of thermodynamics of black hole, one can use the geometrothermodynamical formalism.
\\
\\
{\bf Acknowledgements}\\
We are grateful to the anonymous referee for the insightful comments and suggestions, which have allowed us to improve this paper significantly.
SNS and NR acknowledge the support of Shahid Beheshti University. SHH wishes to thank Shiraz University Research Council. The work of SHH has been supported financially by the Research Institute for Astronomy and Astrophysics of Maragha, Iran.

\end{document}